\PassOptionsToPackage{table,xcdraw}{xcolor}
\documentclass[a4paper,11pt]{article}
\pdfoutput=1

\usepackage{jcappub} 
\usepackage{natbib}
\setcitestyle{square,comma,numbers,sort&compress}
\usepackage{enumerate}
\usepackage{tabularx}
\usepackage[left = 1.0in, top=1.0in,right=1.0in,bottom=1.0in,a4paper]{geometry}
\usepackage[dvipsnames]{xcolor}
\usepackage{comment}
\usepackage[caption=false]{subfig}
\usepackage{cancel}

\usepackage[section]{placeins}
\usepackage{listings}
\usepackage{amsbsy}
\newcolumntype{Y}{>{\centering\arraybackslash}X}

\definecolor{myRED}{rgb}{0.8, 0.25, 0.33}
\usepackage[normalem]{ulem}
\usepackage{dsfont}
\usepackage{pbox}
\usepackage{graphicx}
\usepackage{multirow}
\usepackage{tikz}
\usetikzlibrary{arrows}
\usepackage{enumitem} 
\usepackage{ulem}
\usepackage{soul}
\usepackage{lipsum}  

\usepackage{xcolor,pifont}

\usepackage[many]{tcolorbox}
\definecolor{LightCyan}{rgb}{0.88, 1, 1}
\usepackage{multirow}
\usepackage{booktabs}

\DeclareMathOperator{\sgn}{sgn}

\def\SO{\mathrm{SO}}

\def\MGUT{M_{\mathrm{GUT}}}
\def\MSUSY{M_{\mathrm{SUSY}}}
\def\MEW{M_{\mathrm{EW}}}
\def\OMEGA{\Omega_{\text{DM}}}

\title{\boldmath\huge   
A new perspective on the CMSSM: Yukawa \\ Unification, DM and the SUSY scale}

\author[a]{Stefan Antusch,}
\author[b]{Shaikh Saad,}
\author[c]{and Vasja Susič}

\affiliation[a]{Department of Physics, University of Basel, Klingelbergstrasse\ 82, CH-4056 Basel, \\Switzerland}

\affiliation[b]{Jožef Stefan Institute, Jamova 39, P.~O.~Box 3000, SI-1001 Ljubljana, Slovenia}

\affiliation[c]{Istituto Nazionale di Fisica Nucleare, Laboratori Nazionali di Frascati, \\
C.P.~13, 00044 Frascati, Italy}

\emailAdd{stefan.antusch@unibas.ch, shaikh.saad@ijs.si, vasja.susic@lnf.infn.it}

\abstract{
What does third family ($t$-$b$-$\tau$) Yukawa unification, a typical prediction from embedding the Standard Model (SM) fermions in 16-plets of a grand unifying $\mathrm{SO}(10)$ gauge symmetry, imply for the scale of the supersymmetric (SUSY) partners? Which neutralino dark matter candidate can be realized, and how large is the dark matter relic density? In this work, we address these questions in a simplified SUSY-breaking framework: the Constrained Minimal Supersymmetric Standard Model (CMSSM). To this end, we recast the parameter space of the CMSSM in a way that for all parameter points the SM-like Higgs mass is correctly reproduced. Considering fixed $\tan\beta$ and $\sgn(\mu)$, for every point in the $(x := \frac{M_{1/2}}{m_0}, y :=\frac{A_0}{m_0})$ parameter plane ranges for all observables are predicted. This provides a new perspective on where in parameter space different types of dark matter (DM) are realized, and which value of the SUSY scale is required in order to explain the observed mass of the SM-like Higgs boson. In our analysis we consider and compare two strategies: grid scans over the $(x,y)$ and $(x,y,\tan\beta)$ parameter regions (with iterative RG evolution), as well as MCMC sampling. We find both techniques yield similar results. For $t$-$b$-$\tau$ unification within $5\,\%$ or $10\,\%$, we find $\mu<0$, the SUSY spectrum showing a characteristic pattern, and the SUSY scale around $\mathcal{O}(10)\,\mathrm{TeV}$.
The extra MSSM Higgses are the lowest lying new states at $\sim 2\div 3\,\mathrm{TeV}$ (with discovery potential at the HL-LHC), the $\mathcal{O}(10)\,\mathrm{TeV}$ stops and gluino are in reach of a possible FCC-hh, while bino DM has a mass above $2.5\,\mathrm{TeV}$, is overabundant, and effectively unobservable in planned direct and indirect detection experiments. The DM relic density requires a dilution factor of $10<\mathcal{D}<1000$, implying non-standard cosmology that could leave its imprints in the stochastic gravitational wave background.
}
\makeatletter
\gdef\@fpheader{}
\makeatother
\begin{document}
\maketitle
\flushbottom

\section{Introduction}

Despite intense searches at the LHC, no supersymmetric (SUSY) partners of the Standard Model (SM) particles have been discovered so far. With no new states in the immediate proximity of the electroweak (EW) scale, and 
the default expectations for the SUSY spectrum based on it solving the hierarchy problem~\cite{Weinberg:1975gm,Gildener:1976ai,Susskind:1978ms,tHooft:1980xss} apparently unrealized, this raises the question at which masses do we actually expect to find the SUSY partners. For providing examples of predictive scenarios, Grand Unified Theories (GUTs)~\cite{Pati:1973rp,Pati:1974yy, Georgi:1974sy, Georgi:1974yf, Georgi:1974my, Fritzsch:1974nn} are an interesting framework. Since they unify the SM fermions in joint GUT representations, they strongly restrict the free parameters in the soft SUSY-breaking sector, and furthermore typically imply restrictions for the Yukawa couplings at the GUT scale. In such a reduced parameter space, and taking the observed mass of the SM-like Higgs boson as a constraint, predictions for the SUSY particle spectrum can emerge, as discussed e.g.~in   \cite{Antusch:2015nwi,Antusch:2017ano,Antusch:2019gmc}. A typical consequence from unifying the Standard Model fermions in 16-plets of a grand unifying SO(10) gauge symmetry, can be third family ($t$-$b$-$\tau$) Yukawa unification, which we will focus on in this paper.

In particular, to arrive at an example for a highly predictive (although somewhat simplified) scenario, we consider the Constrained Minimal Supersymmetric Standard Model (CMSSM)~\cite{Drees:1992am,Kane:1993td}.   
The CMSSM is a parametrization of soft SUSY breaking within the Minimal Supersymmetric Standard Model (MSSM), in which the soft terms are universal at some high energy scale, usually associated to the Planck scale (i.e.~mSUGRA~\cite{Drees:1992am}) or the scale of Grand Unification. 

The parameter space of the CMSSM has been extensively studied in the past 2 decades, see e.g.~\cite{Kane:1993td,Ellis:1999mm,Ellis:2001msa,Chattopadhyay:2003xi,Baer:2003yh,Ellis:2003cw,Lahanas:2003yz,Ellis:2010kf,Cao:2011sn,Ellis:2012aa,Ellis:2012nv,Buchmueller:2013psa,Buchmueller:2015uqa,Ellis:2015rya,Bagnaschi:2018igf,Ellis:2018jyl,Ellis:2019fwf}. A recent comprehensive study \cite{Ellis:2022emx} confronts its shrinking parameter space with a host of updated experimental constraints: these include bounds set on the SUSY spectrum by the LHC, as well as constraints on dark matter (DM) from direct and indirect detection measurements, concluding that completely viable patches still exist. One particular assumption, however, which is often present also in other studies involving CMSSM and DM, is that the predicted DM relic density $\OMEGA^{*}\,h^{2}$ calculated assuming standard cosmological evolution
should directly match the one observed experimentally: \hbox{$\OMEGA^{0}\,h^2= 0.12\pm 0.001$} from measurements of the Cosmic Microwave Background (CMB) by the Planck satellite~\cite{Planck:2018vyg}, where $h=0.67$ encodes the present-day expansion rate of the Universe. This, however, tacitly assumes the standard calculation took all the DM details correctly into account. A predicted underabundance \hbox{$\OMEGA^{*} < \OMEGA^{0}$} could indicate, for example, that the MSSM neutralino is only a fraction of DM and the majority component needs to be sought in an extension of the theory. Conversely, a predicted overabundance \hbox{$\OMEGA^* > \OMEGA^0$} could be resolved by subsequent dilution arising from a scenario of non-standard cosmology, such as late-time entropy production through the decay of a super-weakly coupled particle. To this end, the ratio (see, e.g., Ref.~\cite{Hasenkamp:2010if})
\begin{align}
    \mathcal{D} := \OMEGA^{*}/\OMEGA^{0}
    \label{eq:dilution-factor-definition}
\end{align}
defines the required \textit{dilution factor} after freeze-out for a consistent description of the MSSM neutralino being the DM. Such dilution is often present in local SUSY scenarios, when the SUSY breaking is taken into account. The sgoldstino, which gives mass to the gravitino in the super-Higgs mechanism, couples only gravitationally and thus typically decays late and produces entropy that dilutes a previously generated DM abundance.

Our goal in this paper is to study the feasibility of $t$-$b$-$\tau$ Yukawa unification within CMSSM; the analysis does not fit the DM relic abundance, but instead treats the dilution factor $\mathcal{D}$ as a prediction. Other predicted observables we compute are the cross-sections for direct and indirect detection, as well as the SUSY spectrum. Regarding the latter, special attention is given to the masses of extra MSSM Higgs states, which have been shown recently to be generically lighter in $t$-$b$-$\tau$ unified scenarios than the superpartner particles~\cite{Antusch:2019gmc}. 

The above considerations result in a scenario not covered by existing comprehensive studies of the CMSSM parameter space, or complementarily by the existing $t$-$b$-$\tau$ unification literature~\cite{Baer:1999mc,Baer:2000jj,Baer:2001yy,Blazek:2001sb,Blazek:2002ta,Auto:2003ys,Baer:2008xc,Baer:2008jn,Altmannshofer:2008vr,Gogoladze:2010fu,Badziak:2011wm,Gogoladze:2011aa,Gogoladze:2011ce,Anandakrishnan:2012tj,Badziak:2012mm,Baer:2012cp,Baer:2012jp,Joshipura:2012sr,Karagiannakis:2012sv,Ajaib:2013zha,Anandakrishnan:2013cwa,Badziak:2013eda,Anandakrishnan:2014nea,Shafi:2015lfa,Altin:2017sxx,Hussain:2018xiy,Antusch:2019gmc,Gomez:2020gav,Hicyilmaz:2021onw,Ahmed:2022ibc,Shafi:2023sqa}. Indeed, some studies that have addressed Yukawa unification are from before the Higgs mass measurement~\cite{Baer:1999mc,Baer:2000jj,Baer:2001yy,Blazek:2001sb,Blazek:2002ta,Auto:2003ys,Baer:2008xc,Baer:2008jn,Altmannshofer:2008vr,Gogoladze:2010fu,Badziak:2011wm,Gogoladze:2011aa,Gogoladze:2011ce}, while later ones cover boundary conditions different from CMSSM (see Section~\ref{sec:tbtau} for a more detailed discussion). Although we considered both signs of $\mu$ in our study, the usually dominant gluino-loop contribution to the SUSY threshold corrections only goes in the right direction for $\mu<0$, cf.~e.g.~\cite{Hall:1993gn,Antusch:2008tf},  
so it was not surprising for us that we only find viable parameter space for this sign choice.

The physics motivation behind such a scenario is imagined to be a $\SO(10)$ GUT, where 3rd family Yukawa unification originates from the operator $\mathcal{O}=16_{F}\cdot 16_{F}\cdot 10$. Since this operator by itself cannot explain the full richness of the Yukawa parameters at the EW scale, other Yukawa operators (renormalizable or non-renormalizable) must be involved. It is thus sensible to assume that the operator $\mathcal{O}$ merely dominates the 3rd family contribution rather than being the only source, implying only approximate $t$-$b$-$\tau$ unification. This same relaxation of Yukawa unification also helps phenomenologically by raising the extra Higgs masses above the LHC bounds, see~\cite{Antusch:2019gmc}.

Finally, the CMSSM parametrization does not quite match the $\SO(10)$ boundary conditions at the GUT scale, in particular the soft masses for the representations with the Higgs and sfermions have no reason to be the same. 
CMSSM also assumes, beyond SO(10), flavor universality. Although it mainly serves as a further simplifying assumption, we note that it may be justified to some extent by additional flavor symmetries that are often added to explain, e.g., the charged fermion mass hierarchies and the observed quark and lepton flavor mixings.
We use the CMSSM boundary conditions as a simplifying assumption for the analysis in this paper motivated by two reasons:
\begin{itemize}
    \item[(i)] The CMSSM parameter space is smaller, and thus more tractable. A higher-dimensional parameter space might need to be sampled by a method such as Markov Chain Monte Carlo (MCMC), where the interpretation of results requires some care. The CMSSM, however, allows also for a systematic scan in a dense grid, enabling a comparison of the two methods, which can be used as a springboard for an MCMC analysis of a larger parameter space.
    \item[(ii)] To showcase a useful reparametrization of the boundary conditions on a simple example. In CMSSM, one typically discusses the parameter space in terms of a few 2D slices, e.g.~in the $(m_{0},M_{1/2})$ plane at a few fixed values of $A_{0}$ as in \cite{Ellis:2022emx}, or in terms of point clouds projected to 2D scatter plots. From these methods, it is sometimes hard to develop an understanding of the underlying physics. Our analysis uses a reparametrization of the standard CMSSM set, which we find to be physically more illuminating, and could be extended to more complicated examples.
\end{itemize}

We organize the paper as follows: we highlight the virtues of the reparametrization of the usual CMSSM parameters in Section~\ref{sec:reparametrization}, perform our main analysis of approximate $t$-$b$-$\tau$ Yukawa-unified regions in Section~\ref{sec:tbtau}, and then conclude in Section~\ref{sec:conclusions}.


\section{A fresh look at the CMSSM \label{sec:reparametrization}}

The CP-preserving\footnote{While $m_{0}$ and $M_{1/2}$ can be taken real without loss of generality, CP-conservation in the CMSSM refers to real $\mu$ and $A_0$~\cite{Demir:1999ky}.} 
CMSSM~\cite{Kane:1993td} is based on the parameters,
\begin{align}
    m_0, && M_{1/2},  && A_0,  && \tan\beta,  && \sgn(\mu), \label{eq:CMSSM-usual-parameters}
\end{align}
which are used to specify the SUSY soft breaking sector at a high scale (here taken as the Grand Unification scale $\MGUT$). The first four are continuous parameters, while the last one takes the values $\pm 1$. The massive parameters $m_{0}$, $M_{1/2}$ and $A_{0}$ represent the universal sfermion mass, the universal gaugino mass, and the universal proportionality factor between trilinear scalar couplings and the Yukawas: $\mathbf{T}_{x}=A_{0}\,\mathbf{Y}_{x}$ for trilinear coupling matrices $\mathbf{T}$ and Yukawa matrices $\mathbf{Y}$, for all sectors $x=u,d,e$. 

Typically, the four-dimensional parameter space is discussed in terms of two-dimensional slices, often in the $(m_0,M_{1/2})$ plane. Experimental constraints or measurements, e.g.~the mass of the SM-like Higgs, then favor or discard regions in such slices. Clearly, just based on such considerations, it is difficult to get a full picture.

The idea, to which the ``fresh look'' moniker refers to, is based on the introduction of the following set of high scale parameters:
\begin{align}
    x &:= \frac{M_{1/2}}{m_0}, & 
    y &:=\frac{A_0}{m_0}, & 
    m_0, && 
    \tan\beta, &&  
    \sgn(\mu).
    \label{eq:CMSSM-new-parameters}
\end{align}
This reparametrization retains $m_{0}$ as the only massive parameter, and is motivated by the following observations: 
\begin{enumerate}
    \item 
        The parameter $m_{0}$ has almost direct control of the SUSY scale $Q$ (defined as the geometric mean of two stop masses) and the SUSY spectrum more broadly. 
        \par
        More specifically, 
        under an overall rescaling\footnote{Each massive parameter rescales according to their mass dimension, i.e.~squared-mass parameters rescale with $m_{0}^2$.} by a factor $m_{0}$ of all MSSM soft parameters of interest
        ---the trilinear scalar terms $\mathbf{T}_{x}$, the soft masses $m^2_{x}$ and the gaugino masses $M_{x}$---all terms in their renormalization group equations (RGEs) also rescale with the same
        overall factor, cf.~e.g.~\cite{Antusch:2015nwi}. The RGEs namely do not involve any other massive parameter, such as the coupling $\mu$ (of the $H_{u}H_{d}$ term in the superpotential) or the associated soft parameter $b$. This implies that for fixed $x$ and $y$, the RG-run values at a given renormalization scale $\mu_{r}$ of all the listed soft parameters are determined up to an overall factor, namely an appropriate power of $m_{0}$. 
        \par
        With respect to computing the SUSY scale $Q$, the dominant effect of $m_{0}$ is to simply set the scale, while the change in the amount of running needed (and the consequent shift in ratios of soft parameters) is a subdominant log-like effect, resulting in a larger $Q$ for larger $m_{0}$. 
        \par 
        Setting the SUSY scale $Q$, we can then compute the EW vacuum and the SUSY spectrum. For the vacuum, we solve the stationarity conditions for $\mu$ and $b$, which at tree level take the following form (see e.g.~\cite{Martin:1997ns}):
        \begin{align}
            |\mu|^2 &= 
                \frac{1}{2} \left|\frac{m^{2}_{H_{d}}-m^{2}_{H_u}}{\cos(2\beta)}\right|
                -\frac{m^{2}_{H_u}+m^{2}_{H_d}+m_{Z}^{2}}{2}, \label{eq:vacuum-mu}\\
            b&= \frac{m^{2}_{H_{u}}+m^{2}_{H_{d}}+2|\mu|^2}{2}\,\sin(2\beta), \label{eq:vacuum-b}
        \end{align}
        where the soft parameters $m^{2}_{H_{u}}$ and $m^{2}_{H_d}$ are taken at the scale $Q$, while $m_{Z}$ is the $Z$-mass. If $|\mu|\gg m_{Z}$, then the constant term $m_{Z}^2$ in Eq.~\eqref{eq:vacuum-mu} is small, and the dominant effect of rescaling $m_{0}$ is an overall factor in $\mu$, and consequently due to Eq.~\eqref{eq:vacuum-b} also in $b$. For large SUSY scales, i.e.~$Q\gg v$ and the (non-rescaling with $m_{0}$) EW VEVs have a negligible effect in the SUSY spectrum. This results in the SUSY spectrum also merely rescaling by a $m_{0}$ factor to a good approximation. An analysis using ratios $x$ and $y$ thus constrains the dominant rescaling effects into a single parameter $m_{0}$. 
    \item
        The mass of the SM Higgs
        \begin{equation}
             m_h\big(x,y,m_0,\tan\beta,\sgn(\mu)\big) \label{eq:Higgs-mass-function}
        \end{equation}
        increases (for most parameter values) monotonously with $m_0$ when the other parameters in the new set are fixed. The implication is that for fixed values of $\tan \beta$ and $\sgn(\mu)$, the CMSSM slices in the $(x= \frac{m_{1/2}}{m_0}, y=\frac{A_0}{m_0})$ parameter plane then provide a comprehensive picture, since the parameter $m_0$ is set by the measured value $m_h \approx 125\,\mathrm{GeV}$. The effective elimination of one parameter due to the Higgs mass constraint clearly offers an advantage in understanding the parameter space. 
        \par
        The claim that $m_{h}$ increases with $m_{0}$ in the new parametrization can be understood intuitively. The relevant regime for the SUSY scale $Q$ in our analysis and consistent with LHC SUSY bounds is above $\sim\mathrm{TeV}$, requiring an EFT calculation of $m_{h}$,
        see e.g.~\cite{PardoVega:2015eno,Bagnaschi:2017xid,Bagnaschi:2019esc}.
        There is a known tendency that increasing $Q$ increases the Higgs quartic coupling $\lambda$ at the EW scale, and hence the Higgs mass
        $m_{h}$. The main disruption to this tendency arises from the threshold corrections to $\lambda$ from stop mixing, depending directly on the parameter
        \begin{align}
            \hat{X}_{t}& := \frac{A_{t}-\mu/\tan\beta}{Q},
        \end{align}
        cf.~Eq.~(88) in~\cite{Draper:2016pys}. As discussed in the previous point, the new parametrization offers an improved correlation of $m_{0}$ with all massive SUSY parameters, the dominant effect becoming a simple rescaling. An increase in $m_{0}$ with other values in Eq.~\eqref{eq:CMSSM-new-parameters} thus causes a rescaling increase in $Q$, but keeps the stop mixing parameter $\hat{X}_{t}$ constant (unlike what happens in the usual CMSSM parametrization), implying rather robustly an increase in $m_{h}$ in the EFT calculation. 
        \par
        Since there are many other terms in the threshold corrections to $m_{h}$ beside the one depending on $\hat{X}_{t}$, it is impractical to make an analytic justification more detailed than the above for every part of parameter space. Providing further justification for the claim on $m_{h}$ increasing with $Q$, however, we do observe this tendency in our scans of parameter space empirically using the Higgs spectrum tools \texttt{FeynHiggs} and \texttt{SusyHD}, see Section~\ref{sec:computational-setup}; barring numerical instabilities, the tendency holds for all checked points in our grid scans of Sections~\ref{sec:grid-example} and \ref{sec:tbtau-approaches}. 
\end{enumerate}

Further advantages of the new parametrization become obvious when considering the obtained results from exploring the parameter space. To this end we provide an instructive example in Section~\ref{sec:grid-example}.

\subsection{Computational setup \label{sec:computational-setup}}

Before proceeding to the example exploring the $(x,y)$ plane, we state here our general computational setup, relevant for all analyses in the paper. The GUT-scale inputs are given at $\MGUT = 2 \cdot 10^{16}\,\mathrm{GeV}$, and they consist of the following $10$ values:
\begin{align}
    m_{0}, &&
    x=M_{1/2}/m_{0}, &&
    y=A_{0}/m_{0}, &&
    \tan\beta, &&
    y_{t}, &&
    y_{b}, &&
    y_{\tau}, &&
    g_{1}, &&
    g_{2}, &&
    g_{3}, \label{eq:input-parameters}
\end{align}
as well as a discrete choice of $\sgn\mu=\pm 1$, where $y_{t,b,\tau}$ are the 3rd family MSSM Yukawa couplings and $g_{1,2,3}$ are the MSSM gauge couplings. We then perform a top-down computation from the GUT scale to the EW scale using the following pipeline:
\begin{enumerate}
    \item 
        The RGE running is performed by \texttt{SusyTCProton}~\cite{Antusch:2020ztu,Antusch:2015nwi}, which is based on the \texttt{REAP}~\cite{Antusch:2005gp} framework. The package runs in the softly-broken MSSM from $\MGUT$ down to the SUSY scale $\MSUSY \equiv Q$, which is determined during run-time as the geometric mean of the two running masses of the stop states. At $\MSUSY$, the software computes the 1-loop SUSY spectrum\footnote{We use an upgraded version of \texttt{SusyTCProton}, not yet made public, which computes the SUSY spectrum at 1-loop rather than tree-level, extending the expressions from \cite{Pierce:1996zz,Haber:1984rc}
        to include full inter-generational mixing.
        } and matches to the SM RGEs, then runs down to the EW scale---identified with the mass of the $Z$-boson: $\MEW\equiv M_{Z} = 91.1876\,\mathrm{GeV}$ \cite{ParticleDataGroup:2022pth}.  Crucially, the matching of the Yukawa couplings at $\MSUSY$ includes SUSY threshold corrections, cf.~\cite{Antusch:2015nwi}. All RGEs are at $2$-loop order.
    \item 
        The spectrum of the Higgs sector is obtained from \texttt{FeynHiggs 2.19.0}~\cite{Bahl:2017aev,Bahl:2016brp,Hahn:2013ria,Frank:2006yh,Degrassi:2002fi,Heinemeyer:1998np,Heinemeyer:1998yj}. In particular, we use the EFT-like settings \texttt{looplevel=0} and \texttt{loglevel=3} suitable for high SUSY scales above a few $\mathrm{TeV}$, cf.~e.g.~\cite{Bahl:2017aev,Bahl:2019hmm}.
        The masses of the extra Higgses are computed at $1$-loop level. Alternatively, one can compute only the SM Higgs mass using \texttt{SusyHD}~\cite{PardoVega:2015eno}, which uses exclusively the EFT approach, and gives results analogous to \texttt{FeynHiggs}---it predicts a Higgs mass for relevant points about $0.5\,\mathrm{GeV}$ higher. 
        Both \texttt{FeynHiggs} and \texttt{SusyHD}
        also report estimated errors, which are in the $0.5\div 0.7\,\mathrm{GeV}$ range, implying consistent results. In our analysis, we make use of the \texttt{FeynHiggs} results for the SM and extra Higgs masses.
    \item 
        DM related quantities are obtained via \texttt{MicrOMEGAs 5.2.7a}~\cite{Belanger:2020gnr,Belanger:2004yn,Belanger:2001fz,Belanger:2018ccd,Barducci:2016pcb,Belanger:2014vza,Belanger:2013oya,Belanger:2010gh,Belanger:2008sj,Belanger:2006is}. 
        Specifically, the extracted quantities are the DM relic abundance $\OMEGA^*$ (computed using standard cosmology), the neutralino-nucleon (direct detection) and the neutralino-neutralino (indirect detection) cross-sections. In our analysis, we apply the recent astrophysical bounds on both spin-independent and spin-dependent WIMP–nucleon cross sections. The strongest direct detection (DD) limits come from the \textit{LUX-ZEPLIN} (LZ) experiment~\cite{LZ:2024zvo}, which improve upon earlier results from \textit{XENONnT}~\cite{XENON:2023cxc}, previous LZ measurements~\cite{LZ:2022lsv}, and \textit{PandaX-4T}~\cite{PandaX-4T:2021bab,PandaX:2024qfu}. It is worth mentioning that the current LZ result provides the strongest SI exclusion on the DM–nucleon scattering cross section, setting a limit of $2\times 10^{-47}$ cm$^2$ at 90\% confidence level (CL) for a DM mass of order a TeV. For completeness, we also project the future sensitivity of \textit{DARWIN}~\cite{DARWIN:2016hyl} and show the irreducible neutrino background~\cite{Billard:2013qya} (from coherent neutrino scattering~\cite{Freedman:1973yd}). Moreover, WIMPs are expected to annihilate at the present time, leading to the possibility of detecting their annihilation products—most notably, gamma rays, which currently represent the most promising indirect detection (ID) channel for dark matter. Therefore, we also impose ID bounds from gamma-ray observations. For WIMPs in the MSSM framework, the most relevant ID constraints arise from dark matter annihilation channels such as $\chi\chi \to W^+W^-$, $ZZ$, and $\tau^+ \tau^-$. The gamma-ray detectors that currently provide the most stringent limits for these channels include the space-based \textit{Fermi-LAT}~\cite{Fermi-LAT:2015att} and \textit{AMS-02}~\cite{Cuoco:2016eej} instruments, as well as the ground-based telescopes \textit{MAGIC}~\cite{MAGIC:2016xys} and \textit{H.E.S.S.}~\cite{HESS:2022ygk}. At dark matter masses around the $\mathrm{TeV}$ scale---relevant to our study---these experiments set strong upper bounds on the annihilation cross section, typically of the order $\langle \sigma v \rangle \lesssim 10^{-26} \, \mathrm{cm}^3/\mathrm{s}$.  
    \item
        Finally, to quantify the compatibility of predictions with EW-scale experimental data, we evaluate the $\chi^{2}$ function for the low energy observables given in Table~\ref{tab:chi2-observables}. These consist of the 3rd family Yukawas, the gauge couplings, and the SM Higgs mass~\cite{Antusch:fit}.  Conversely, the DM relic abundance $\OMEGA^*$, or rather the dilution factor $\mathcal{D}$ from Eq.~\eqref{eq:dilution-factor-definition}, is treated as a prediction and is not included as an observable into $\chi^2$.
        \def\SKIP{2pt}
        \begin{table}[htb]
        \centering
        \begin{tabular}{llll}
        \toprule
        observable&value $y$&$\pm\delta y^{(\text{exp})}$&$\pm  \delta y^{(\chi^{2})}$\\
        \midrule
        $y_t$ & 
            $0.986$ & ${}_{-0.004}^{+0.004}$ & ${}_{-0.00986}^{+0.00986}$ \\[\SKIP]
        $y_{b}/10^{-2}$ & 
	       $1.63$ & ${}_{-0.01}^{+0.02}$ & ${}_{-0.0163}^{+0.02}$ \\[\SKIP]
        $y_{\tau}/10^{-3}$ & 
	       $9.9370$ & ${}_{-0.0014}^{+0.0015}$ & ${}_{-0.09937}^{+0.09937}$ \\[\SKIP]
        $g_{1}$ & 
	       $0.461227$ & ${}_{-0.000027}^{+0.000025}$ & ${}_{-0.00230614}^{+0.00230614}$ \\[\SKIP]
        $g_{2}$ & 
	       $0.650963$ & ${}_{-0.000037}^{+0.000035}$ & ${}_{-0.00325482}^{+0.00325482}$ \\[\SKIP]
        $g_{3}$ & 
	       $1.21183$ & ${}_{-0.00467}^{+0.00448}$ & ${}_{-0.00605915}^{+0.00605915}$ \\[\SKIP]
        $m_{h}\; [\mathrm{GeV}]$ & 
            $125.09$ & $0.24$ & $\sim 0.52^\star$\\
        \bottomrule
    \end{tabular}
    \caption{
        The list of observables in the $\chi^{2}$, their values $y$ and experimental errors $\pm\delta y^{(\text{exp})}$ at the scale $\MEW$, as well as the errors $\pm\delta y^{(\chi^{2})}$ taken in the $\chi^{2}$-function. The latter are based on experimental errors, but taken to be at least $1\,\%$ for Yukawa couplings and $0.5\,\%$ for gauge couplings to prevent over-fitting beyond the theoretical error of next-order contributions in perturbation theory. The error in the Higgs mass is taken to be the theoretical error reported by \texttt{FeynHiggs}, with the $\star$ next to its typical value.
        The input values of the gauge couplings and the quark and lepton masses are taken from Ref.~\cite{Antusch:fit}, which updates the quark masses and mixings previously presented in Refs.~\cite{Antusch:2013jca}. The Higgs mass $m_h= 125.09\pm 0.24\,\mathrm{GeV}$ comes from \cite{ATLAS:2015yey}. 
        \label{tab:chi2-observables}
    }
    \end{table}
        \par
        In order to exclude points whose predictions violate phenomenological bounds, a sufficiently high penalization is added to the $\chi^{2}$ in such cases. We consider the following for penalization: direct- and indirect-detection bounds for DM (see previous point), and experimental bounds from searches for additional Higgs bosons. The latter are the most important constraints for the $t$-$b$-$\tau$ unified scenario. Specifically, the heavy neutral Higgs bosons with masses $m_{\phi} > m_h$ ($\phi = H, A$) have been searched for at the LHC. Currently, the most stringent bounds come from the production of a single neutral Higgs, which subsequently decays via $\phi \to \tau^+ \tau^-$. Since 3rd family Yukawa unification, even if approximate, requires $\tan\beta$ to be large, the couplings of $\phi$ to down-type quarks and charged leptons are enhanced. Therefore, gluon fusion leads to $gg \to \phi$ and $b\overline{b}\phi$ (where $b$ denotes a bottom quark), whereas the scattering process leads to $gb \to b\phi$. Due to the null observation of any such signals in a data sample collected with the CMS~\cite{CMS:2022goy} as well as ATLAS~\cite{ATLAS:2020zms} detectors at $\sqrt{s} = 13~\mathrm{TeV}$, corresponding to an integrated luminosity of 138~$\mathrm{fb}^{-1}$, stringent limits are placed on the mass of the neutral component of the heavy Higgs bosons. For example, with $\tan\beta \to 60$, LHC searches translate to $m_\phi \gtrsim 2~\mathrm{TeV}$ at $95\,\%$ C.L.~(see Fig.~\ref{fig:mcmc-Higgs}).
\end{enumerate}

In summary, the above pipeline provides the computational framework from the initial input parameters to observables. 
Compatibility with measured low energy observables is quantified by the $\chi^{2}$ function, with an additional penalization if experimental bounds of as-of-yet unobserved phenomena are violated. Observables not considered in either, such as the dilution factor $\mathcal{D}$ and the SUSY spectrum, are also stored for each point. The described pipeline is a tool used for exploring the parameter space, either on a grid, cf.~Sections~\ref{sec:grid-example} and \ref{sec:tbtau-approaches}, or when sampling via MCMC, cf.~Section~\ref{sec:tbtau-approaches}.

We conclude with a few comments on the general setup of inputs and outputs:
\begin{itemize}[topsep=0.1cm,itemsep=0.0cm]
    \item The rationale for including the 3rd family Yukawa couplings among the input parameters of Eq.~\eqref{eq:input-parameters} is that they have a sizable impact on RG-running of soft SUSY-breaking parameters. The price to pay is that they then need to be included among the observables in Table~\ref{tab:chi2-observables}. This is in contrast to 1st and 2nd generation Yukawas, which have little numeric impact on either the RGEs of the soft parameters or the SUSY spectrum, and as such can be safely neglected from the analysis (their values can be taken zero or a small fixed value for the purposes of \texttt{SusyTCProton}).
    \item Our rationale for considering unified boundary conditions for soft parameters (simplified to CMSSM) and $t$-$b$-$\tau$ Yukawa unification is a suppositions of an underlying $\SO(10)$ GUT symmetry. This, however, requires unification of MSSM gauge couplings at $
    \MGUT$, so the computational setup where $g_{i}(\MGUT)$ are free inputs would seem at odds with having a single unified coupling value. There are two things to note here. First, we remain agnostic about the full $\SO(10)$ theory at $\MGUT$ and in particular the scalar sector. If the masses of these fields are scattered an order of magnitude around $\MGUT$, the associated threshold corrections to the gauge couplings can easily be at the level of a few $\%$. Under these mild assumptions, only approximate (at the $\%$ level) and not exact gauge coupling unification in MSSM is required. Second, as is well known, low scale SUSY unifies MSSM gauge couplings around the scale $
    \MGUT\approx 2\cdot 10^{16}\,\mathrm{GeV}$ (the reason we take this value in the first place). This feature carries over to the SUSY scales predicted in our analysis, in particular all MSSM gauge coupling inputs at $\MGUT$ automatically come out in the range $0.67\div 0.7$, so they unify within a few $\%$ in accordance with the requirements from threshold corrections.
\end{itemize}

\subsection{The CMSSM parameter space revisited: the $(x,y)$ parameter region \label{sec:grid-example}}

As we now discuss, looking at the CMSSM in the $(x,y)$-plane offers a new perspective that allows to obtain an improved understanding of the implications and predictions from certain regions of parameter space, regarding e.g.~dark matter and the viability of GUT scale predictions for Yukawa ratios such as third family Yukawa unification motivated by $\SO(10)$ GUTs.

As an example, we consider a parameter scan in the $(x= \frac{m_{1/2}}{m_0}, y=\frac{A_0}{m_0})$ plane for $\tan\beta=52$ and $\sgn(\mu)$ negative (motivated by our later focus on Yukawa unification), following the strategy described in the main part of Section~\ref{sec:reparametrization} such that all the shown parameter points feature the measured SM-like Higgs mass $m_h \approx 125$ GeV. More specifically, for a fixed point in the $(x,y)$-plane (and fixed values $\tan\beta$ and $\sgn\mu$), the $\chi^{2}$ function of Section~\ref{sec:computational-setup} is fit to be near zero---effectively setting the correct inputs $\{g_{i},y_{x},m_{0}\}$ just below $\MGUT$ to get the correct $\{g_{i},y_{x},m_{h}\}$ at $\MEW$ (a fit of $7$ parameters to $7$ observables, cf.~Eq.~\eqref{eq:input-parameters} and Table~\ref{tab:chi2-observables}).  

\begin{figure}
    \centering
    \includegraphics[width=0.48\linewidth]{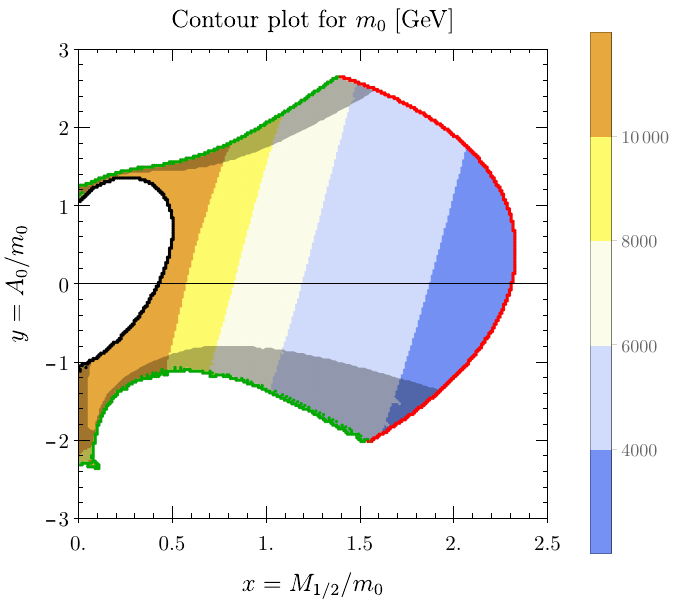}
    \hspace{0.02\linewidth}
    \includegraphics[width=0.48\linewidth]{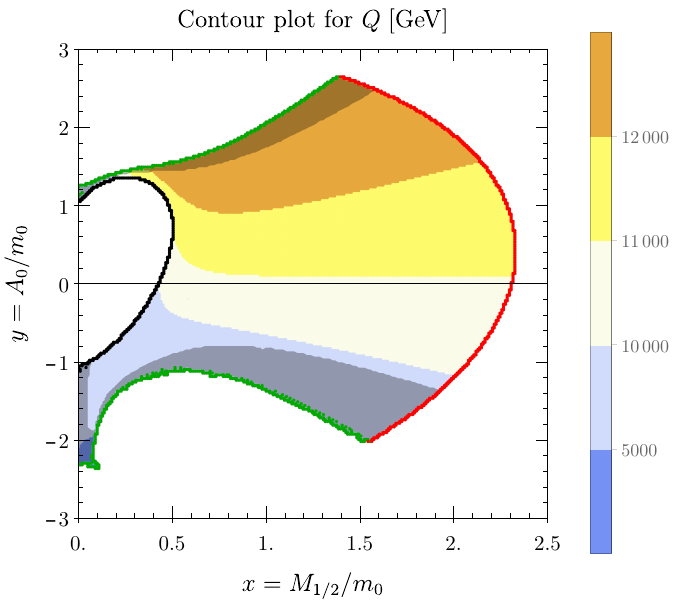}\\
    \includegraphics[width=0.48\linewidth]{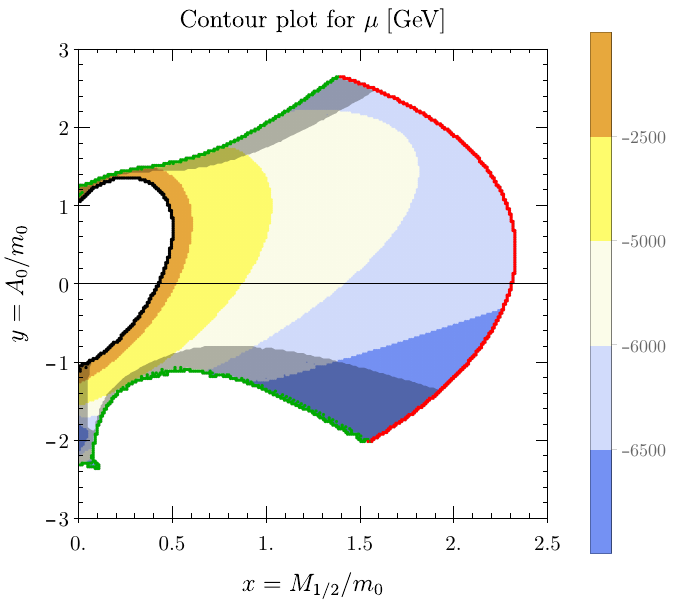}
    \hspace{0.02\linewidth}
    \includegraphics[width=0.48\linewidth]{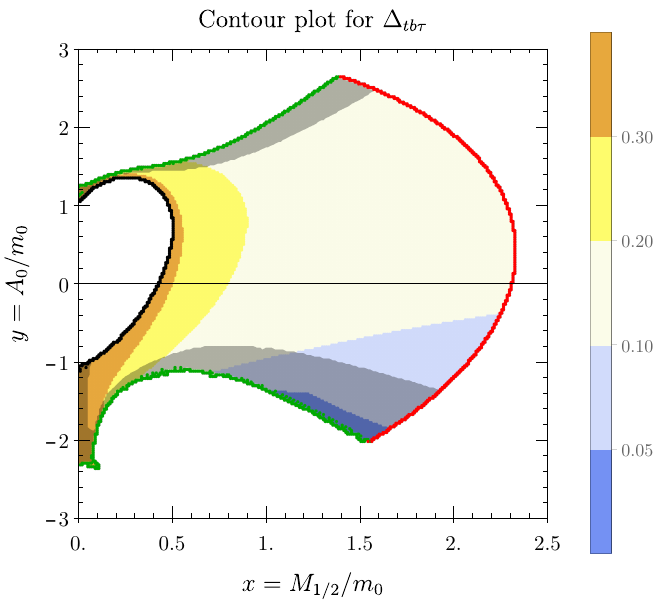}\\
    \includegraphics[width=0.48\linewidth]{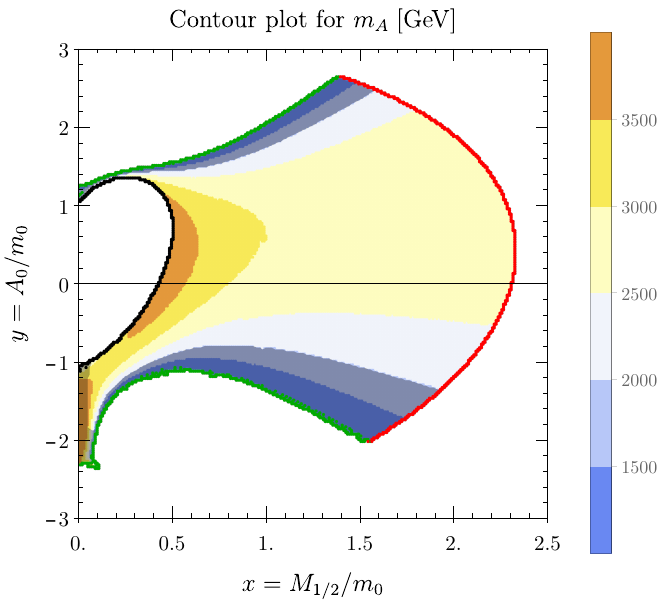}
    \hspace{0.02\linewidth}
    \includegraphics[width=0.48\linewidth]{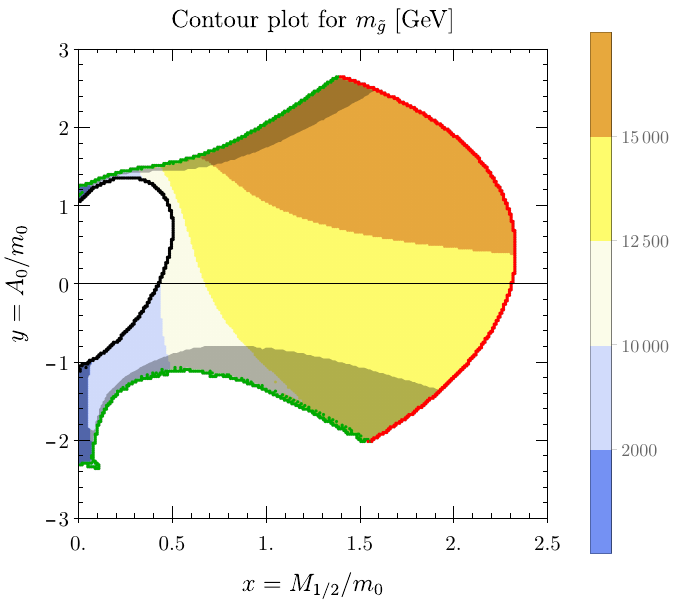}\\
    \caption{
        Results from an example parameter scan in the $(x= \frac{M_{1/2}}{m_0}, y=\frac{A_0}{m_0})$ plane for $\tan\beta=52$ and $\sgn(\mu)=-1$, following the strategy where $m_{0}$ is determined by requiring the SM Higgs mass prediction $m_{h}\simeq 125.1\,\mathrm{GeV}$. The region of viable points is unshaded and is enclosed by boundaries characterized by different modes of failure: no EW vacuum (black), neutralino is not LSP (red), extra Higgses/gluino mass in conflict with experimental bounds (shaded, then green).
        Each panel represents contours for a different quantity of interest: the CMSSM parameter $m_{0}$, the SUSY scale $Q$, the value of the MSSM superpotential parameter $\mu$ relevant for EW symmetry breaking, the Yukawa non-unification parameter $\Delta_{tb\tau}$ defined in Eq.~\eqref{eq:definition-Yukawa-nonunification}, the extra Higgs mass $m_{A}$ and the gluino mass $m_{\tilde{g}}$. See main text for further details and discussion.
    }
    \label{fig:grid2d-main}
\end{figure}

The results in the $(x,y)$-plane are shown in Figures~\ref{fig:grid2d-main} and \ref{fig:grid2d-DM}. They are based on a grid with a resolution $(\delta_x,\delta_y)\sim (0.010,0.023)$. Descriptions of the plots are given in the captions, while some further observations are made below:
\begin{itemize}[leftmargin=1.0cm,itemsep=0.0cm]
    \item 
        The region where the neutralino is the lightest supersymmetric particle (LSP) is bounded in the $(x,y)$-plane, allowing for a comprehensive scan and compact description of the parameter space. The boundaries are colored in accordance with the reason the point is no longer viable:
        \begin{itemize}[leftmargin=0.5cm,itemsep=0.0cm]
            \item At the \textbf{red boundary}, the neutralino is no longer the LSP.
            \item On the upper and lower \textbf{green boundary}, the extra Higgs masses of the MSSM are too close the the SM Higgs mass, so that \texttt{FeynHiggs} is not reporting a result. 
            \item The \textbf{black boundary} signifies an ellipse, in which EW symmetry breaking fails, as discussed in more detail below.
            \item The \textbf{shaded grey region} represent points which are phenomenologically excluded; in particular, the predicted masses $m_{A}$ or $m_{\tilde{g}}$ of the extra Higgs or the gluino, respectively, are below the experimental bounds set by the LHC (see Sec.~\eqref{sec:computational-setup} and Fig.~\ref{fig:mcmc-Higgs} for extra Higgses, while a rough cut \hbox{$m_{\tilde{g}}>2\,\mathrm{TeV}$} is made for gluinos~\cite{ATLAS:2022ihe}). The extra Higgs bound in particular supersedes the green boundary to delimit the viable parameter region, making the details of the exact failure of \texttt{FeynHiggs} at the green boundary extraneous. 
        \end{itemize}
    \item 
        The upper pair of plots in Figure~\ref{fig:grid2d-main} show the predicted values of the CMSSM parameter $m_{0}$ and the SUSY scale $Q$ which result in the correct SM Higgs mass. We see that contour lines have different angles in the two plots, an effect of RGE running and the stop masses not depending only on soft masses $m^{2}_{\tilde{Q}_{3}}$ and $m^{2}_{\tilde{u}^{c}_{3}}$---the stop mixing depends on the trilinear parameter $A_{t}$, which is most directly controlled by the new parameter $y$. Furthermore, the SUSY scale is predicted to be around $10\,\mathrm{TeV}$ in most of the viable region; note that this comes out automatically, i.e.~there is no hidden assumption on the proximity of the SUSY scale to the EW scale.
    \item 
        The elliptically-shaped exclusion region beyond the black boundary is where the computation fails to yield a viable EW symmetry breaking for any $m_{0}$. In detail, the 1-loop analog of Eq.~\eqref{eq:vacuum-mu} predicts $|\mu|^{2}<0$. In the viable region close to the black ellipse, $|\mu|$ should thus be predicted to be small (close to zero), which is indeed confirmed by the contours of $\mu$ in the middle-left panel of Figure~\ref{fig:grid2d-main}. The region where $|\mu|$ is predicted to be rather small compared to the values of the usual CMSSM parameters $m_{0}$, $M_{1/2}$ and $A_{0}$ is known in the literature as the focus-point region, cf.~e.g.~\cite{Feng:1999mn,Baer:2005ky,Feng:2011aa,Draper:2013cka}, which is in the $(x,y)$ parametrization mapped close to the ellipse. 
    \item 
        With a high value $\tan\beta=52$, we have the possibility of $t$-$b$-$\tau$ Yukawa unification, the subject of our focus in this work, see Section~\ref{sec:tbtau}. Anticipating that analysis, it is interesting to see in which $(x,y)$-regions such approximate 3rd family Yukawa unification can occur. To that end we define the quantity
        \begin{align}
            \Delta_{tb\tau}:= \max_{i,j\in\{t,b,\tau\}} \left[ \frac{y_i(\MGUT)}{y_j(\MGUT)} - 1\right],
            \label{eq:definition-Yukawa-nonunification}
        \end{align}
        which measures the spread of 3rd family MSSM Yukawa coupling parameters at the GUT scale.\footnote{
           The literature on 3rd family Yukawa unification sometimes uses its own measure $R$ (labeled also $R_{tb\tau}$) of approximate unification, introduced in 
           \cite{Auto:2003ys}. It is related to our measure via $R=1+\Delta_{tb\tau}$.
        } 
        We see from the middle-right panel in Figure~\ref{fig:grid2d-main} that Yukawa unification to under $10\,\%$ is possible in the bottom-right part of the viable region, whereas $5\,\%$ is too close to the green boundary and hence excluded by bounds on the extra Higgs mass. Note that $t$-$b$-$\tau$ unification is possible to $5\,\%$ or better for $\tan\beta>52$, see Section~\ref{sec:tbtau}.
        \item 
        In the bottom panels of Figure~\ref{fig:grid2d-main}, contours for the extra Higgs mass $m_{A}$ and gluino mass $m_{\tilde{g}}$ are presented. These demonstrate the origin of the shaded regions: the extra Higgs bounds (of around $\sim 2\,\mathrm{TeV}$) excluded the upper and lower regions adjacent to the green boundary, while a low gluino mass prediction excludes regions with very small $x$. Although the gluino mass bound from the LHC is estimated at $2.4\,\mathrm{TeV}$~\cite{ATLAS:2022ihe,Constantin:2025mex} (derived for some specific scenarios), we shaded the region with a more conservative bound of $2\,\mathrm{TeV}$.  
        \begin{figure}
            \centering
            \includegraphics[width=0.48\linewidth]{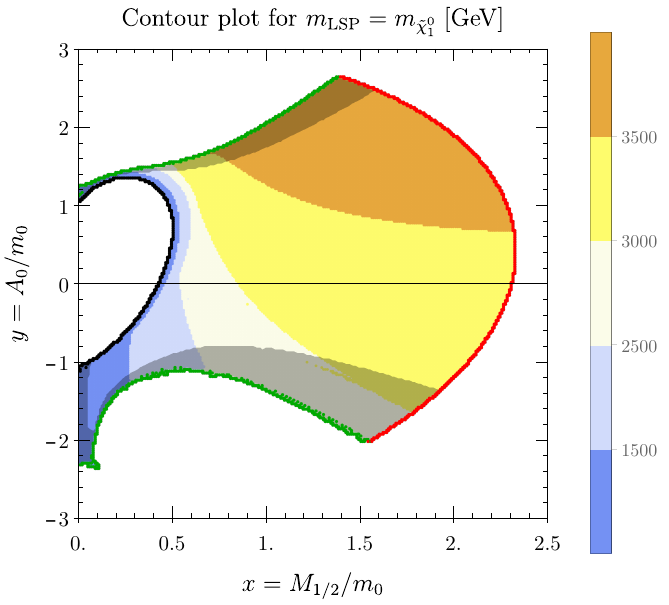}
            \hspace{0.02\linewidth}
            \includegraphics[width=0.48\linewidth]{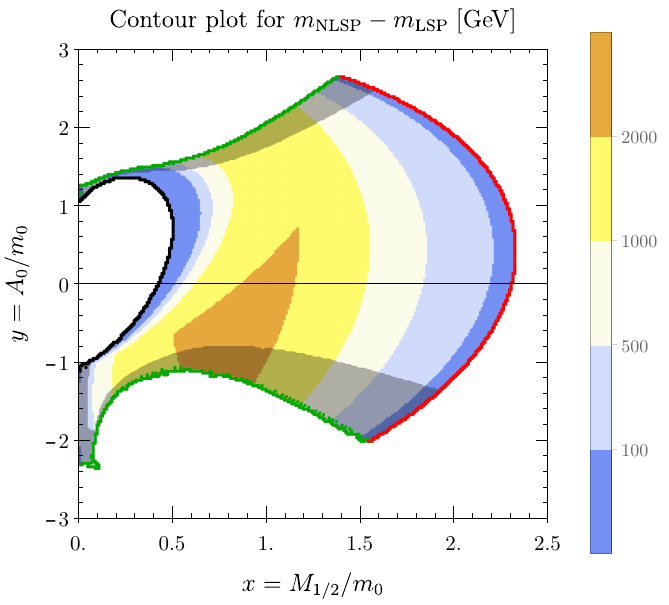}\\
            \includegraphics[width=0.48\linewidth]{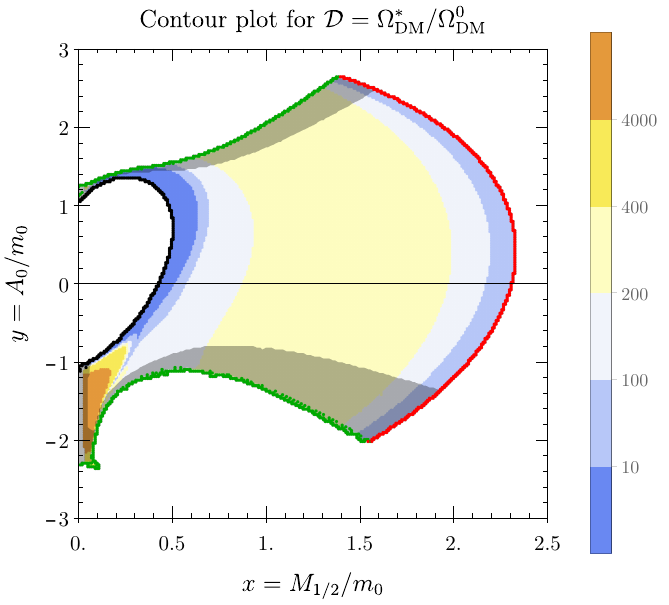}
            \hspace{0.02\linewidth}
            \includegraphics[width=0.48\linewidth]{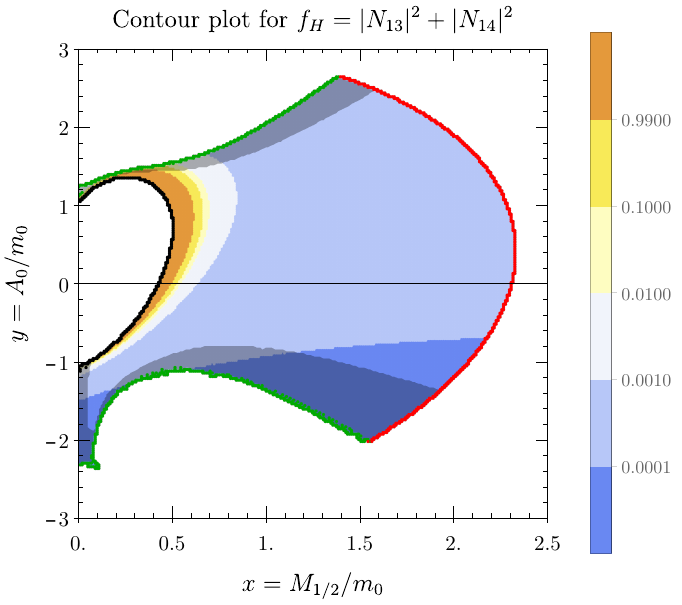}
                \caption{
                    Results from an example parameter scan in the $(x= \frac{m_{1/2}}{m_0}, y=\frac{A_0}{m_0})$ plane for $\tan\beta=52$ and $\sgn(\mu)=-1$, as described in the caption of Figure~\ref{fig:grid2d-main} and the main text. The four panels show contour plots for DM-related quantities: the mass of the neutralino LSP, the mass of the next-lightest SUSY particle $m_{\text{NLSP}}$ (excluding extra Higgses), the predicted dilution factor $\mathcal{D}$ for DM from Eq.~\eqref{eq:dilution-factor-definition}, and the higgsino admixture $f_{H}$ in the LSP, see main text.
                }
                \label{fig:grid2d-DM}
        \end{figure}
        \item 
        The various panels of Figure~\ref{fig:grid2d-DM} are dedicated to understanding the LSP and DM properties. The LSP (lightest neutralino) mass $m_{\text{DM}}\equiv m_{\tilde{\chi}^{0}_{1}}$ is shown in the upper-left panel, and we can see that masses below $1.5\,\mathrm{TeV}$ are reached only near the black boundary. As discussed earlier, proximity to the excluded ellipse region implies $|\mu|$ being close to zero, i.e.~the focus-point region, and thus the LSP neutralino is mostly a higgsino. This identification is supported by the bottom-right panel, where the higgsino admixture portion~\cite{Kowalska:2018toh} $f_{H}=|N_{13}|^{2}+|N_{14}|^{2}$ is shown, where $N_{1i}$ are the relevant entries of the neutralino mixing matrix.  The LSP near the ellipse has automatic co-annihilation partners in another neutral and a charged higgsino, and for a mass $m_{\chi^{0}_{1}}\approx 1.1\,\mathrm{TeV}$~\cite{Arkani-Hamed:2006wnf} the predicted DM relic density $\OMEGA^*$  matches the currently observed one, i.e.~close to the black ellipse there is in fact a region with a dilution factor $\mathcal{D}=1$ consistent with standard cosmology. 
        \par
        Another interesting region is near the red boundary, beyond which the neutralino is no longer the LSP. Approaching it, the next-lightest supersymmetric particle (NLSP) can by definition of the boundary come arbitrarily close in mass to the LSP, cf.~also top-right panel of Figure~\ref{fig:grid2d-DM}. This provides an efficient co-annihilation partner (the stau $\tilde{\tau}$), as confirmed by the region with a moderate DM dilution factor $\mathcal{D}<100$ in the bottom-left panel. The LSP mass is too high, however, to reach $\mathcal{D}<10$ and a prediction for $\OMEGA^*$ consistent with standard cosmology. Other than near the black ellipse boundary, the LSP is identified as the bino, confirmed by the small higgsino admixture in the bottom-right panel. The bino region near the LSP boundary with a moderate dilution factor for DM is thus another region of general interest, and it includes the bottom corner where approximate $t$-$b$-$\tau$ unification can occur, cf.~bottom-right panel of Figure~\ref{fig:grid2d-main}.  
\end{itemize}

We see from the $\tan\beta=52$ example that our approach yields a bounded viable region of characteristic shape in the $(x,y)$ plane, whose features have been well understood. 

\section{$t$-$b$-$\tau$ Yukawa unification and bino DM in the CMSSM \label{sec:tbtau}}

We now turn specifically to (approximate) $t$-$b$-$\tau$ unification in the context of CMSSM. We investigate its parameter space and the implications for DM observables.  We present the results using the new CMSSM parametrization advocated in Section~\ref{sec:grid-example}, and the quantity $\Delta_{tb\tau}$ defined in Eq.~\eqref{eq:definition-Yukawa-nonunification} as a measure of exactness for 3rd family Yukawa unification. An exact relation $y_{t}=y_{b}=y_{\tau}$ in the MSSM at the GUT scale yields $\Delta_{tb\tau}=0$; the regions of interest will be those within $5\,\%$ and $10\,\%$ deviation from Yukawa unification, i.e.~$\Delta_{tb\tau} < 0.05$ and $\Delta_{tb\tau} < 0.10$, respectively.

Although $t$-$b$-$\tau$ unification is well-trodden terrain in the literature~\cite{Baer:1999mc,Baer:2000jj,Baer:2001yy,Blazek:2001sb,Blazek:2002ta,Auto:2003ys,Baer:2008xc,Baer:2008jn,Altmannshofer:2008vr,Gogoladze:2010fu,Badziak:2011wm,Gogoladze:2011aa,Gogoladze:2011ce,Anandakrishnan:2012tj,Badziak:2012mm,Baer:2012cp,Baer:2012jp,Joshipura:2012sr,Karagiannakis:2012sv,Ajaib:2013zha,Anandakrishnan:2013cwa,Badziak:2013eda,Anandakrishnan:2014nea,Shafi:2015lfa,Altin:2017sxx,Hussain:2018xiy,Antusch:2019gmc,Gomez:2020gav,Hicyilmaz:2021onw,Ahmed:2022ibc,Shafi:2023sqa}, remarkably, a study within CMSSM has not been performed (with the exception of \cite{Karagiannakis:2012sv}, where their parametrization of approximate Yukawa unification allows for a solution with $\mu>0$, but with $\Delta_{tb\tau}>1.7$ far outside the range traditionally counted as Yukawa unification). Considering only studies after the measurement of the Higgs mass, the main deviations from the CMSSM constraints is a split $m^{2}_{H_{u}}\neq m^{2}_{H_{d}}$ beyond the minimal $\SO(10)$-like boundary conditions~\cite{Anandakrishnan:2012tj,Badziak:2012mm,Joshipura:2012sr,Anandakrishnan:2013cwa,Badziak:2013eda,Anandakrishnan:2014nea,Altin:2017sxx,
Gomez:2020gav,Ahmed:2022ibc,Shafi:2023sqa} (to facilitate a solution to EW symmetry breaking, especially for $\mu>0$), or non-universal gaugino masses~\cite{Badziak:2012mm,Ajaib:2013zha,Anandakrishnan:2013cwa,Badziak:2013eda,Anandakrishnan:2014nea,Shafi:2015lfa,Altin:2017sxx,Hussain:2018xiy,Gomez:2020gav,Ahmed:2022ibc,Shafi:2023sqa} (such that correct signs in the threshold correction $\delta y_b/y_b$ and the anomalous magnetic dipole moment of the muon $\Delta a_{\mu}^{\text{SUSY}}>0$ are simultaneously possible assuming $\mu M_{3}<0$ and $\mu M_{2}>0$). Furthermore, 
the majority of these studies exclude regions of parameter space where DM relic density $\OMEGA^*$ is computed as overabundant.

Numerical minimization\footnote{
    We use a stochastic version of the differential evolution algorithm~\cite{storn1997differential} (version \texttt{DE/rand/1/bin} with randomly selected $F\in (0.5, 1)$ for every point, cf.~e.g.~\cite{price2005differential}).} 
of the $\chi^{2}$ function we described in Section~\ref{sec:computational-setup} easily yields $\chi^{2}\ll 1$ for $\mu<0$, while the $\mu>0$ case is discarded due to a poor fit with data ($\chi^{2}\gtrsim 150$ when $\Delta_{tb\tau}<10\,\%$). We thus find solutions only for $\mu<0$ (the case where the gluino-loop threshold correction has the correct sign for $y_{b}$).
In what follows, we study the CMSSM parameter space with $\mu<0$ and approximate $t$-$b$-$\tau$ unification in two approaches: a grid scan and MCMC, see Section~\ref{sec:tbtau-approaches}. The results are then presented in Section~\ref{sec:tbtau-results}, while a discussion in Section~\ref{sec:tbtau-DM-overabundance} comments on the results of the DM relic abundance.

\subsection{Two approaches to exploring parameter space\label{sec:tbtau-approaches}}

The parameter space is explored in two different ways: a systematic 3D grid scan and MCMC sampling. The details are as follows:

\begin{enumerate}[topsep=0.1cm,parsep=0.2cm,itemsep=0.2cm]
    \item 
    The systematic grid scan is performed in the parameters $(x,y,\tan\beta)$, with the discrete choice $\sgn(\mu)=-1$. An analogous approach to the 2D grid scan example in the $(x,y)$ space from Section~\ref{sec:grid-example} is used: for a fixed grid value in $(x,y,\tan\beta)$, a minimization of the $\chi^{2}$ function is performed, where the GUT input parameters are listed in Eq.~\ref{eq:input-parameters} (the inputs $\{g_{i},y_{x},m_{0}\}$ are varied), while the observables at the EW scale are listed in Table~\ref{tab:chi2-observables}; see Section~\ref{sec:computational-setup} for further details. Crucially, $m_{0}$ is effectively determined by setting a target for the SM Higgs mass $m_{h}$, see discussion in the main part of Section~\ref{sec:reparametrization}.
    The DM observables computed via \texttt{MicrOmegas} are not part of the $\chi^{2}$, but the results are stored.
    \par
    Note that the minimization of $\chi^{2}$ in every point essentially reaches the value $\chi^{2}=0$, implying an exact fit to the observables. The minimization for a given grid point can be sped up by leveraging minimization results from neighboring points.
    To loosen the rigid constraint of an exact fit, we vary the target SM Higgs mass via $m_{h}=125.1\,\mathrm{GeV}+\delta m_{h}$; taking discrete values for $\delta m_{h}$ results in multiple datasets,    
    which we ultimately combine. 
    \par
    The specific scan we perform is limited to the region $\Delta_{tb\tau}\leq 0.10$ of approximate $t$-$b$-$\tau$ unification. The result is combined from $3$ datasets obtained by using the Higgs mass variations
    $\delta m_{h}\in\{0,-0.5,+0.5\}\,\mathrm{GeV}$ (this is in line with the reported theoretical uncertainty of \texttt{FeynHiggs} of $\sim 0.5\,\mathrm{GeV}$ for the points). The scans are performed with the grid resolution $(\delta_x,\delta_y,\delta_{\tan\beta})\sim(0.025,0.06,0.25)$, yielding datasets of individual sizes as shown in Table~\ref{tab:grid-dataset-size}. We see that phenomenological bounds on these datasets, with the bound on the extra Higgs mass $m_{A}$ turning out to be the only relevant discriminator, shows a preference for predicting a higher SM Higgs mass. 
        \begin{table}[htb]
            \centering

            +\begin{tabular}{lrrr}
                \toprule
                target $m_{h}$ & $124.6\,\mathrm{GeV}$ & $125.1\,\mathrm{GeV}$ & $125.6\,\mathrm{GeV}$\\
                \midrule
                no pheno constraints & $20474$ & $19794$ & $18801$ \\
                pheno constraints & $1523$ & $4891$ & $7660$ \\
                \bottomrule
            \end{tabular}
            \caption{
                Numbers of points from grid scans in $(x=M_{1/2}/m_{0},y=A_{0}/m_{0},\tan\beta)$ and regions compatible with $t$-$b$-$\tau$ unification within $\Delta_{tb\tau}<0.10$. The datasets are computed for different target SM Higgs masses $m_{h}$. The final dataset is combined from all points consistent with phenomenological constraints (bottom row), most crucially with LHC bounds on the extra Higgs mass $m_{A}$.  
                \label{tab:grid-dataset-size}}
        \end{table}
    \item 
    A less rigid exploration of the parameter space is performed by MCMC, since non-perfect fits $\chi^{2}> 0$ may also be accepted in accordance with the algorithm. The explored parameter space is thus expected to be somewhat enlarged. The $\chi^{2}$ function is again that of Section~\ref{sec:computational-setup}. Unlike the grid, all input parameters of Eq.~\eqref{eq:input-parameters} are varied. We only accept points consistent with phenomenological bounds, implemented as a penalization in $
    \chi^{2}$, cf.~Section~\ref{sec:computational-setup}. Most crucially, we impose the LHC bounds on the extra Higgs mass $m_{A}$.
    \par 
    We make use of a custom implementation of the MCMC algorithm based on adaptive step size~\cite{Roberts01012009}. The chosen prior is a uniform distribution in all input parameters. In particular, we choose it flat in the original CMSSM parameters $M_{1/2}$ and $A_{0}$, while the new parametrization is used for presentation purposes. 
    We perform two MCMC scans based on different thresholds of approximate $t$-$b$-$\tau$ unification: $\Delta_{tb\tau}=0.10$ and $\Delta_{tb\tau}=0.05$. After discarding the points from the initial burn-in phase, the two respective datasets consist of $4.8\cdot 10^5$ and $6.48\cdot 10^5$ points. 
\end{enumerate}

We conclude this section by discussing the use of MCMC as a tool for parameter space exploration. The MCMC algorithm is designed to sample from a posterior density distribution, given a likelihood function and a prior distribution, see e.g.~the Statistics section in \cite{ParticleDataGroup:2024cfk}. Our likelihood function is proportional to $e^{-\chi^{2}/2}$, i.e.~we assume Gaussian errors on observables, while we take a uniform distribution for the prior. The latter choice exhibits agnosticism about the GUT scale values of input parameters and puts all the burden on the likelihood function. Note that the number of input parameters in $\chi^2$, cf.~Eq.~\eqref{eq:input-parameters}, is larger than the number of observables in $\chi^{2}$, cf.~Table~\ref{tab:chi2-observables}. 
This implies flat directions of $\chi^{2}=0$ in parameter space; the points in these directions are all equally likely, and the MCMC thus easily explores this space. Given the larger number of inputs, the $\chi^{2}$ function is not interpreted as a measure for goodness-of-fit, but the MCMC sampler still provides the Bayesian posterior density given the data (or profile-likelihood in the frequentist approach) and thus provides a mapping of viable regions of parameter space given current experimental measurements and constraints. Indeed, MCMC sampling has been used for mapping out the CMSSM specifically from very early on, see e.g.~\cite{Baltz:2004aw,Allanach:2005kz,RuizdeAustri:2006iwb}.

As we shall see in Section~\ref{sec:tbtau-results}, the results of both grid sampling and MCMC sampling are consistent. This is important for validating the choice of our MCMC prior, and shows the algorithm efficiently explored the viable parameter space. While grid sampling of the CMSSM parameters was still tractable due to the small dimension of CMSSM, a choice of more realistic GUT boundary conditions for soft SUSY-breaking parameters would result in a parameter space of higher dimension. In this context, the comparison of both methods gives confidence about the reliability of the MCMC approach for future exploration in those cases (especially for explorations of bounded regions in parameter space).  

\subsection{Scan results \label{sec:tbtau-results}}

The result of the grid and MCMC scans described in Section~\ref{sec:tbtau-approaches} are presented in this section. To facilitate the comparison of the two approaches, the results are always plotted together, with the following color conventions:
\begin{itemize}[topsep=0.1cm,itemsep=0.0cm]
    \item \textit{Grey points} are used for scatter plots of the MCMC data. 
    \item \textit{Red} filled-in contours are used for drawing $2$-$\sigma$ regions of highest posterior density (HPD) of the MCMC data. They are drawn over the scatter plots, thus obstructing most of the scatter plots.\footnote{
    In the context of MCMC, scatter plots are a somewhat dubious way of presenting data, since the extent of the plot depends on the length of the chain (a chain of infinite length and drawing points with finite size would formally fill-up the entire region where $\chi^{2}<\infty$). Furthermore, points are drawn on top of each other, obscuring where the density of points is highest. Nevertheless, the drawn scatter plots can still be used to identify ranges achieved by our specific chains (total length around half a million points for each dataset, see Sec.~\ref{sec:tbtau-approaches}), corresponding to roughly $\chi^{2}< 2 \log(5\cdot 10^5)\approx 26$. The robust observable for MCMC are HPD regions, which we also draw.
    }
    \item \textit{Blue contour outlines} are used for drawing the full range of grid datasets. 
\end{itemize} 
Each type of object comes in two shades: the darker shade corresponds to $\Delta_{tb\tau}< 0.10$ and the lighter one to the more stringent constraint $\Delta_{tb\tau}<0.05$. The latter is obtained as a subset of points from the former dataset; all but one of the $5\,\%$-unification points turn out to correspond to the highest SM Higgs mass target of $m_{h}=126.4\,\mathrm{GeV}$, cf.~Table~\ref{tab:grid-dataset-size}.

The plots are displayed in Figures~\ref{fig:mcmc-regions-with-tanB}--\ref{fig:mcmc-DM2}. We comment and discuss these plots below:
\begin{itemize}[topsep=0.2cm,itemsep=0.2cm,parsep=0.2cm]
    \item 
    \underline{\textit{Displaying the viable $(x,y,\tan\beta)$ parameter space:}}\\[5pt]
        Figures~\ref{fig:mcmc-regions-with-tanB} and \ref{fig:mcmc-regions-xy} are intended to represent the 3D region of points in the space $(x,y,\tan\beta)$ by projecting them to the $3$ planes $(x,\tan\beta)$, $(y,\tan\beta)$ and $(x,y)$.
        In both figures, the MCMC datasets are presented as scatter plots (grey) and $2$-$\sigma$ HPD regions (red), while the grid results are represented by 
        either intervals or contours that enclose all the points in the datasets (blue). In particular, delimiters are used to indicate the range of values of $x$ and $y$ for a given $\tan\beta$ in Fig.~\ref{fig:mcmc-regions-with-tanB}, 
        appearing at discrete values of $\tan\beta$ given the grid nature of the scan.
        \begin{figure}[htb]
            \centering
            \includegraphics[width=0.49\textwidth]{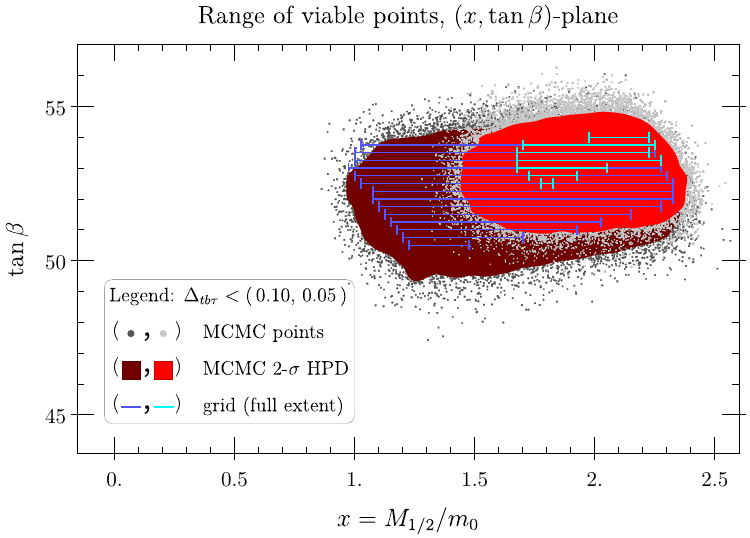}
            \includegraphics[width=0.49\textwidth]{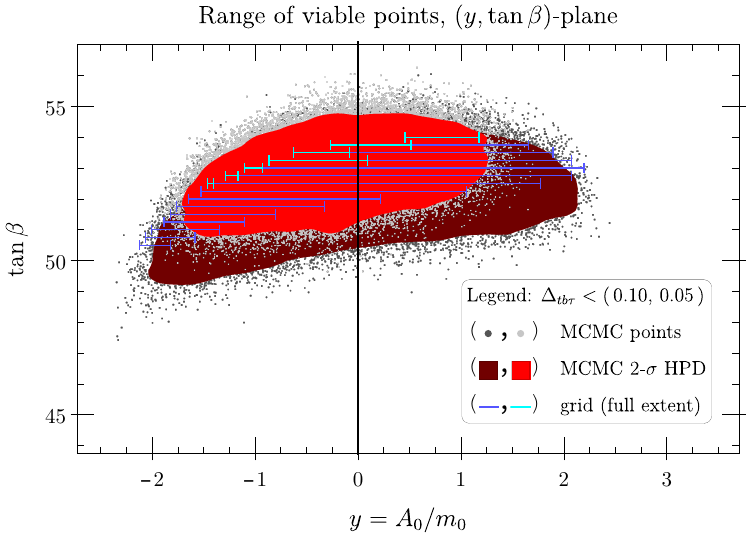}\\
            \caption{
                The range of viable points in the $(x,\tan\beta)$-plane (left) and $(y,\tan\beta)$-plane (right). 
                The grey scatter plots and red $2$-$\sigma$ HPD contours represent MCMC datasets, while the blue delimiters show ranges from grid datasets, see plot legends.  
            \label{fig:mcmc-regions-with-tanB}
            }
        \end{figure}
        \par 
        In line with expectations, the $10\,\%$ non-unification regions (darker shade) present extensions of the $5\,\%$ regions (lighter shade) for every type of plot (grey,red,blue). 
        Comparing the regions predicted by MCMC and the grid, the scatter plot can extend beyond the blue grid contours, since the MCMC does not require an exact fit $\chi^{2}=0$ in the EW observables (gauge and 3rd family Yukawa couplings). The MCMC $2$-$\sigma$ HPD regions can, however, have smaller or larger ranges than the grid contours---the MCMC effect due to $\chi^{2} > 0$ enlarges the region, while taking the $2$-$\sigma$ subset shrinks it.
        \par
        The results in the $(x,y)$-plane in the left panel of Fig.~\ref{fig:mcmc-regions-xy} can be directly compared to the 2D grid results from Section~\ref{sec:grid-example}, see the lowest contour regions of $\Delta_{tb\tau}$ in the bottom-right panel of Figure~\ref{fig:grid2d-main}. The $\Delta_{tb\tau}<0.10$ contour is now enlarged compared to the 2D-grid, since 
        it represents the results from multiple values of $\tan\beta$.
        Interestingly, $\tan\beta > 52$ allows for grid points within $\Delta_{tb\tau} < 0.05$, as can be seen from the appearance of light blue intervals at larger $\tan\beta$ in Fig~\ref{fig:mcmc-regions-with-tanB}.
        \par
        Additionally, the right panel of Fig.~\ref{fig:mcmc-regions-xy} shows the interval ranges for the original MSSM parameters $(m_{0},M_{1/2},A_{0})$. This conveys additional information about the absolute mass scales of parameters for viable points, and facilitates comparison with other CMSSM scans.
        \begin{figure}[htb]
            \centering
            \includegraphics[height=0.4\textwidth]{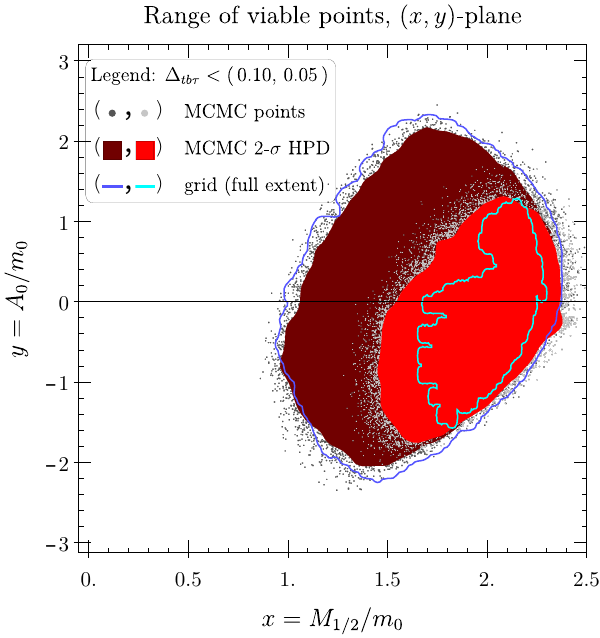}
            \hspace{1cm}
            \includegraphics[height=0.4\textwidth]{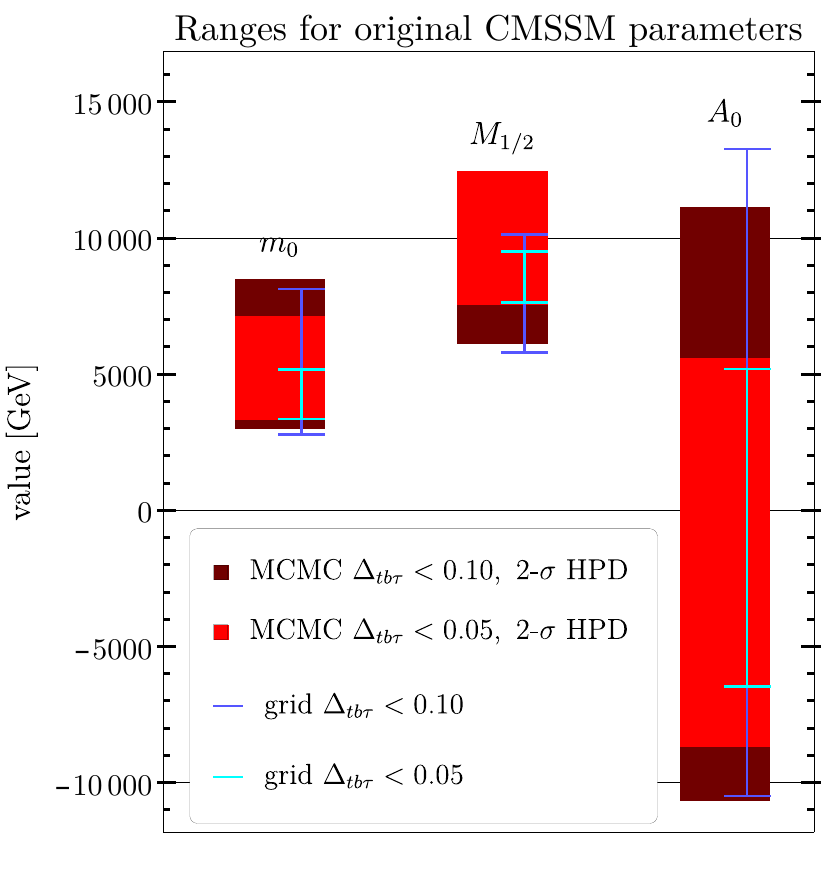}
            \caption{The viable regions in the $(x,y)$-plane (left), and the interval ranges for original CMSSM parameters $(m_{0},M_{1/2},A_{0})$ (right). The grey scatter plots and red $2$-$
            \sigma$ HPD contours/bars represent MCMC datasets, while the blue contours/delimiters show the range from grid datasets, see plot legends.
            } \label{fig:mcmc-regions-xy}
        \end{figure}
    \item 
    \underline{\textit{The Higgs masses and the SUSY scale:}}\\[5pt]
        Two observables of interest are the predicted mass of the SM Higgs $m_{h}$ and of the extra Higgs $m_{A}$, see left and right panel of Fig.~\ref{fig:mcmc-Higgs}, respectively. 
        \begin{figure}[htb]
            \centering
            \includegraphics[width=0.49\textwidth]{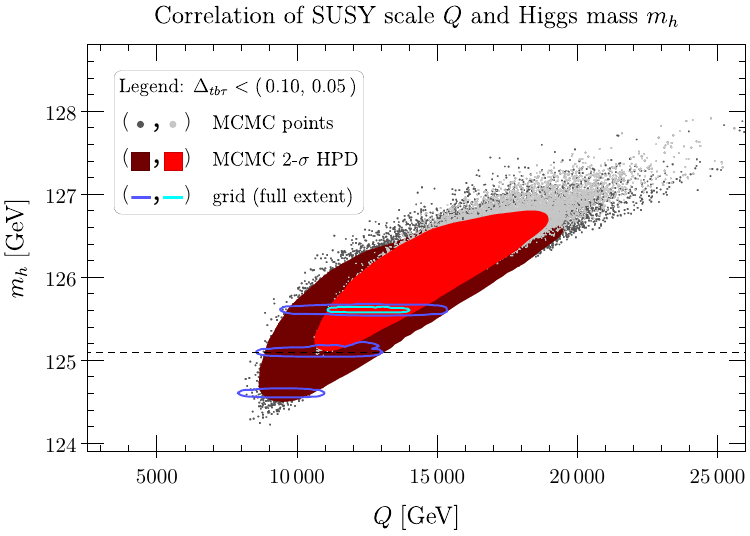}
            \includegraphics[width=0.49\textwidth]{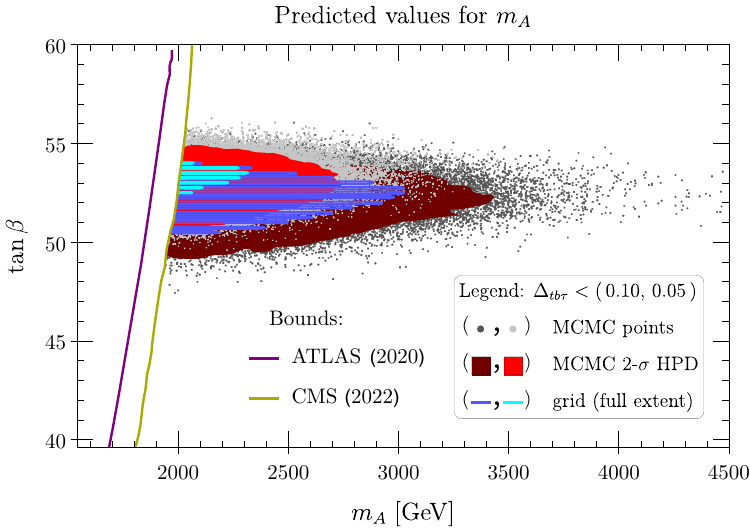}\\
            \caption{
                Results for the MCMC and grid searches. Left panel: the SM Higgs mass $m_{h}$ correlated with the SUSY scale $Q$; there is a clear preference for overshooting rather than undershooting the current experimental bound for $m_{h}$. Right panel: the predicted mass $m_{A}$ of the heavy neutral Higgs;  constraints come from LHC searches for $H/A \to \tau^+ \tau^-$ at $\sqrt{s}=13~\mathrm{TeV}$ with 138~$\mathrm{fb}^{-1}$~\cite{CMS:2022goy,ATLAS:2020zms}, see also Section~\ref{sec:computational-setup}. 
            } \label{fig:mcmc-Higgs}
        \end{figure}
        \par
        The SM Higgs mass is part of the $\chi^{2}$-fit, and thus is predicted with values close to the experimental measurement (horizontal dashed line). The plot in the left panel shows the correlation between the SUSY scale $Q$ and the mass $m_{h}$, i.e.~there is a tendency that a higher SUSY scale also implies a slightly higher SM Higgs mass. The SUSY scale comes out in the range $\sim 9\div 21\,\mathrm{TeV}$. The horizontal blue contours of the grid scan are consistent with the grid dataset using three target values of $m_{h}$. In the MCMC, on the other hand, $m_{h}$ can change, and thus explores an extended range; we see that there is a preference for a larger SUSY scale and larger SM Higgs mass. Notice also that a higher SUSY scale leads to better Yukawa unification, as has been pointed out even before the SM Higgs mass was known, see~Section~3.1 in \cite{Auto:2003ys}.
        \par
        The right panel of Fig.~\ref{fig:mcmc-Higgs} shows the mass of the CP-odd extra Higgs state $A$. As pointed out in \cite{Antusch:2019gmc}, $t$-$b$-$\tau$ unification predicts comparatively low extra Higgs masses, and the plot indeed shows that LHC bounds cut abruptly into the predicted range.\footnote{
            The analysis in \cite{Antusch:2019gmc} differs from the present work in its use of $\SO(10)$ boundary conditions, it does not impose the neutralino LSP or consider DM observables at all, and does not impose detailed bounds on extra Higgs states (which have been significantly improved by the LHC since its publication).
        }
        A larger degree of non-unification (the $10\,\%$ dataset) or less rigid EW constraints (MCMC) both allow for larger values of $m_{A}$; the $2$-$\sigma$ HPDs predicts a range $[2, 2.7]\,\mathrm{TeV}$ for up to $5\,\%$ Yukawa non-unification and $[2,3.4]\,\mathrm{TeV}$ for up to $10\,\%$ unification. If any of these scenarios are realized in nature, finding the extra MSSM Higgs states in that range would most probably be the first experimental indication. Indeed,
        based on an early CMS report~\cite{CMS:2018oxh} for the High-Luminosity LHC (HL-LHC), the sensitivity to $m_{A}$ might rise as high as $\sim 4\,\mathrm{TeV}$ for $\tan\beta=60$. We (naively) estimate this reach from the results for lower $\tan\beta$ in Ref.~\cite{CMS:2018oxh}, which indicate, for example, that an existing limit of about $1.2\,\mathrm{TeV}$ ($1\,\mathrm{TeV}$) would extend to above $2\,\mathrm{TeV}$ at the HL-LHC with an integrated luminosity of $3\,\mathrm{ab}^{-1}$ ($6\,\mathrm{ab}^{-1}$). If this bound indeed applies, the entire parameter space of Fig.~\ref{fig:mcmc-Higgs} will be probed by the HL-LHC. In this context, a dedicated study of the HL-LHC sensitivity in the higher mass range for $m_{A}$ and larger $\tan\beta$ would be highly desirable.
    \item 
    \underline{\textit{The predicted SUSY spectrum:}}\\[5pt]
    An important prediction for our scenario is also the MSSM SUSY spectrum. The results in the form of $2$-$\sigma$ HPD intervals for MCMC datasets (red) and the full predicted range for grid datasets (blue) are shown in Fig.~\ref{fig:mcmc-susy-spectrum}.
    \begin{figure}[htb]
        \centering
        \includegraphics[width=0.95\textwidth]{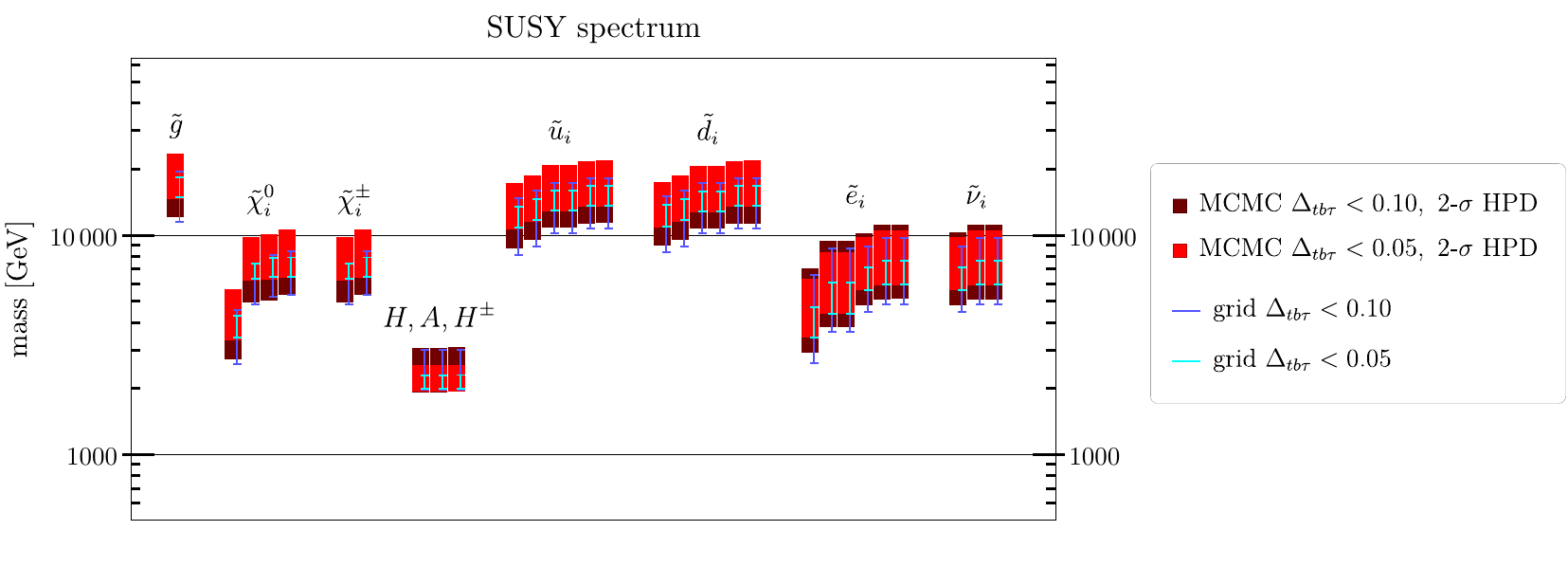}
        \caption{The SUSY spectrum in logarithmic scale of the MCMC ($2$-$\sigma$ HPD, in red) and grid (full range, blue delimiters) datasets. The darker (lighter) shades indicate datasets with better than $10\,\%$ ($5\,\%$) $t$-$b$-$\tau$ unification, as shown in the legend. State labels: $\tilde{g}$ is the gluino, $\tilde{\chi}_{i}^{0}$ are neutralinos, $\tilde{\chi}_{i}^{\pm}$ are charginos, $H,A,H^{\pm}$ are the extra Higgses, while $\tilde{u}_{i}$, $\tilde{d}_{i}$, $\tilde{e}_{i}$ and $\tilde{\nu}_{i}$ are sfermion mass eigenstates.
        } \label{fig:mcmc-susy-spectrum}
    \end{figure}
    \par
    The spectrum overview confirms our earlier observation, that the extra Higgs states are comparatively light; they are in fact the lightest of the new states, and most directly within experimental reach. In terms of superpartner states with negative $R$-parity, the lightest neutralino $\tilde{\chi}^{0}_{1}$ (with the bino being the dominant admixed component) is the LSP, with the lightest charged slepton $\tilde{e}_{1}$ (the stau) being the NLSP; this is important for DM co-annihilation, as discussed further in the next bullet point. The squarks $\tilde{u}_{i}$ and $\tilde{d}_{i}$, as well as the gluino $\tilde{g}$, are predicted mostly above $10\,\mathrm{TeV}$. 
    \par
    In summary, the spectrum is heavy enough to be consistent with LHC bounds, but manifests sufficient proximity to the EW scale that a discovery at a future hadron collider such as FCC-hh is possible. For example, in a scenario where pair-produced stops each dominantly decay to a top quark and the lightest neutralino, namely $\tilde{t}\to t\,\tilde{\chi}^0_{1}$, the full potential of FCC-hh $\sqrt{s} = 100\,\mathrm{TeV}$ with $30\,\mathrm{ab}^{-1}$ luminosity could probe stop masses up to $\sim 10\,\mathrm{TeV}$~\cite{Gouskos:2642475,Golling:2016gvc} for DM mass as heavy as our scenario. As for the gluino, assuming a massless neutralino, the FCC-hh reach can be as high as $17\,\mathrm{TeV}$ if each pair-produced gluino decays to jets and missing transverse energy, $\widetilde{g} \to jj \, \widetilde{\chi}^0$. In scenarios where the gluino is nearly degenerate in mass with the LSP, the missing transverse momentum in the event is reduced, and consequently, the sensitivity decreases to about $7.5\,\mathrm{TeV}$ (where for this analysis, a mass splitting of $10\,\mathrm{GeV}$ is assumed)~\cite{Golling:2016gvc,FCC:2018byv}. In our scenario, gluino is predicted to be much heavier than the LSP, hence the expected reach should be closer to the case of $17\,\mathrm{TeV}$. The quoted bounds suggest a very good chance that the gluino $\tilde{g}$ and the stop $\tilde{t}_{1}$ from our scenario would be within reach of FCC-hh.
    \item 
    \underline{\textit{Predictions for dark matter observables:}}\\[5pt]
    First, the LSP mass $m_{\text{DM}}$ and necessary dilution factor $\mathcal{D}$ for DM relic abundance are shown in Fig.~\ref{fig:mcmc1-DM1}. In the $\Delta_{tb\tau}<0.10$ datasets, the mass $m_{\text{DM}}$ can go as low as $2.5\,\mathrm{TeV}$, while the upper end is around $4.7\,\mathrm{TeV}$ for the grid, $6\,\mathrm{TeV}$ for the bulk of MCMC points ($2$-$\sigma$ HPD), while individual MCMC points can predict masses as high as $\sim 8\,\mathrm{TeV}$. The less rigid predictions for the EW observables in the MCMC can thus clearly extend the upper bounds of the predicted range compared to the grid. Conversely, the low-end range for the mass $m_{\text{DM}}$ and dilution $\mathcal{D}$ shows that only MCMC outliers fill in the full $\Delta_{tb\tau}<10\,\%$ contour, which implies that those predictions originate from a tuned part of parameter space representing a relatively small fraction of the total viable volume for the MCMC. In particular, a smaller predicted relic abundance $\OMEGA^{*}$ is reached when the NLSP is very close in mass to the LSP (implying efficient co-annihilation), as well as when the LSP mass itself is low. 
    \begin{figure}[htb]
            \centering
            \includegraphics[width=0.7\textwidth]{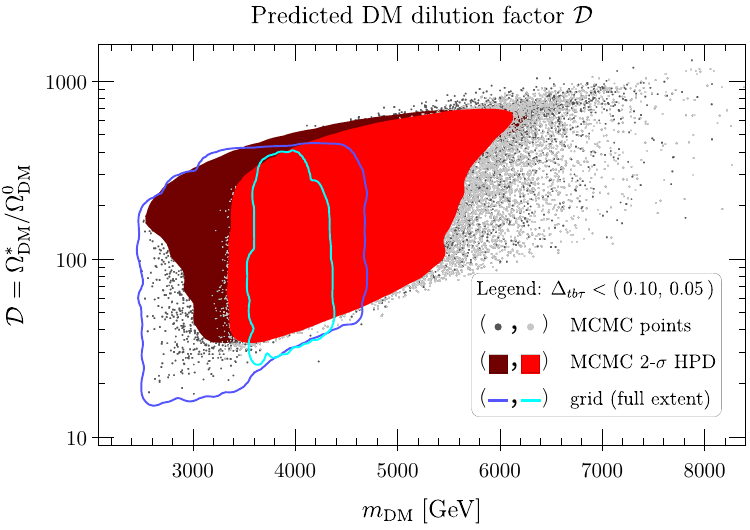}
            \caption{The correlation between the predicted DM mass and dilution factor $\mathcal{D}$ (needed to bring the dark matter relic abundance $\OMEGA^*$ computed assuming standard cosmology down to its currently measured value $\OMEGA^0$). The results compare the MCMC and grid datasets, each shown with two levels of strictness $\Delta_{tb\tau}$ for 3rd family Yukawa unification, see legend.} 
            \label{fig:mcmc1-DM1}
    \end{figure}
    \par
    Specifically regarding the dilution factor, it is predicted to be roughly below $\mathcal{D}<1000$; it can go as low as $\mathcal{D}\approx 18$, with a large part of the $2$-$\sigma$ contours extending below $\mathcal{D}<100$ in general. Albeit the NLSP can be arbitrarily close to the bino LSP (arbitrarily close to the analog of the red boundary from the 2D grid in Section~\ref{sec:grid-example}), the bino mass itself cannot be sufficiently lowered to reach $\mathcal{D}=1$, implying the DM abundance prediction in standard cosmology is in tension with predicting the correct Higgs mass (which depends on $m_{0}$, on which $m_{\text{DM}}$ also predominantly depends on).
    \par
    The observables for bino direct detection (DD) via recoil from nucleons, or indirect detection (ID) from bino-bino annihilation, is shown in Figure~\ref{fig:mcmc-DM2}. The predicted cross-sections are well below bounds of current or future experiments. Furthermore, for DD, the points are also below the threshold of coherent neutrino scattering off nucleons (dashed black curve), i.e.~they are deep in the region of ``neutrino fog''---below this threshold, DD experiments fundamentally lose sensitivity, 
    as potential DM signals become indistinguishable from the background noise generated by neutrinos.
    The only MCMC outliers rising above this threshold are in the case of the spin independent cross section (upper panel). The results thus show that the $t$-$b$-$\tau$ scenario being investigated would not be within experimental reach from the point of view of DM detection experiments. 
    \par
\begin{figure}
    \centering
    \includegraphics[height=0.42\textwidth]{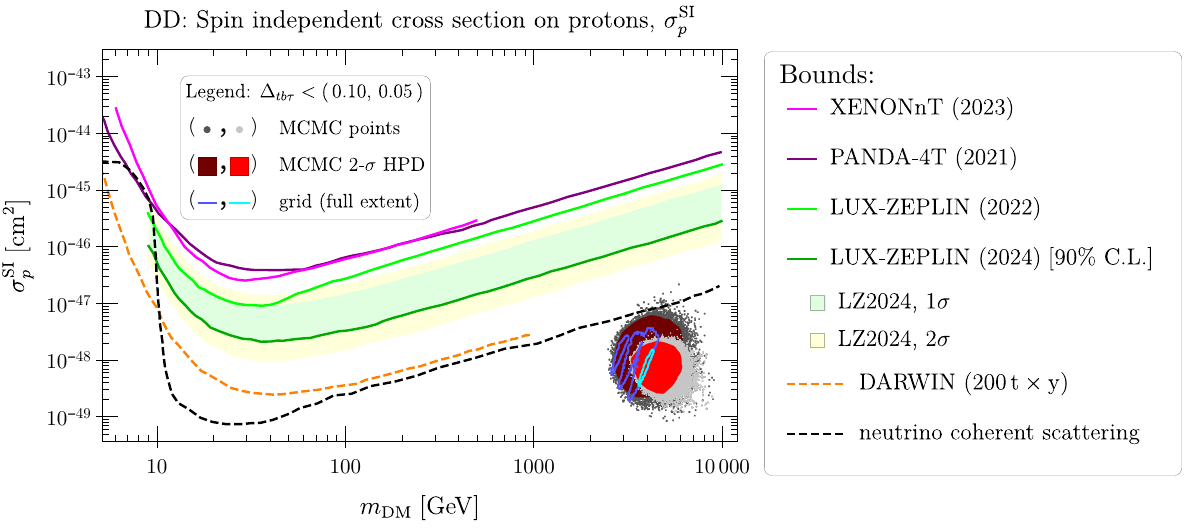}\\[10pt]
    \includegraphics[height=0.42\textwidth]{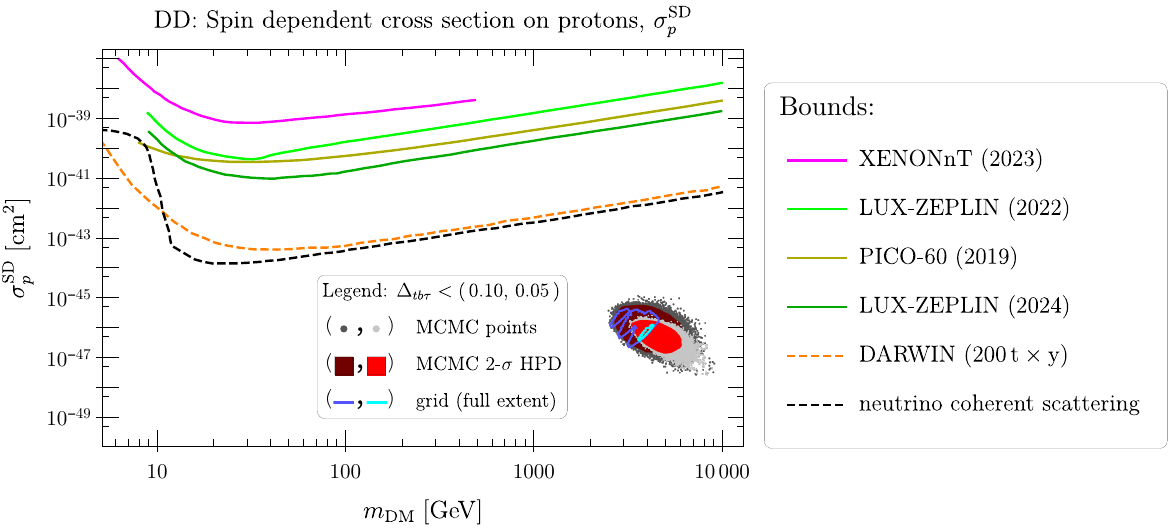}\\[10pt]
    \includegraphics[height=0.42\textwidth]{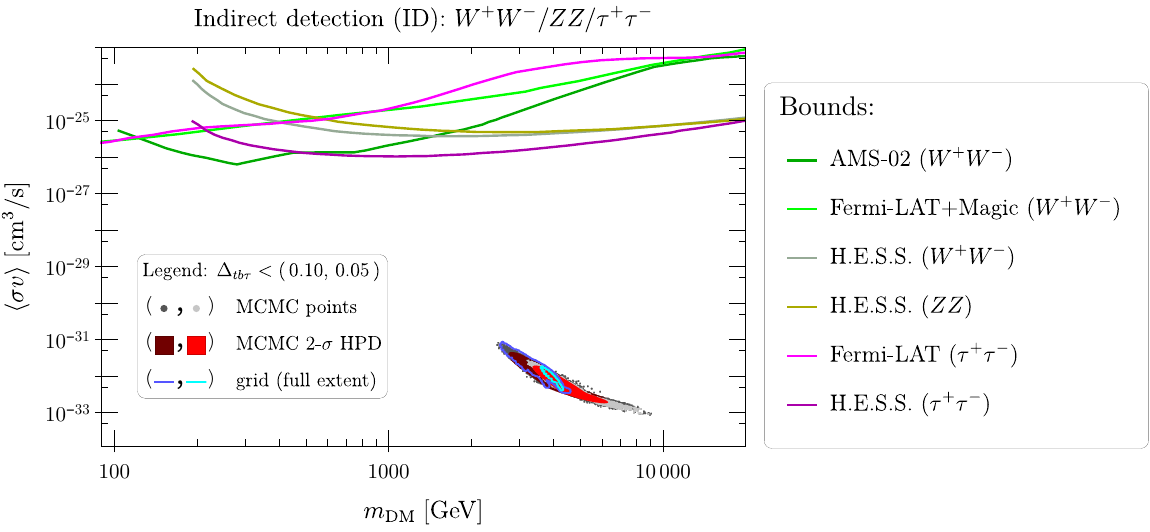}\\
    \caption{The predictions for DM observables for direct and indirect detection (DD and ID, respectively) in our $t$-$b$-$\tau$ unification scenario. The predictions are well below the experimental bounds, both present (solid curves) and future (dashed curves), see main text for more details. For DD, both the spin independent (SI, upper panel) and spin dependent (SD, middle panel) cross sections of DM scattering off protons are shown, labeled $\sigma_{p}^{\text{SI}}$ and $\sigma_{p}^{\text{SD}}$, respectively. Essentially identical plots apply for DM scattering off neutrons, i.e.~for $\sigma_{n}^{\text{SI}}$ and $\sigma_{n}^{\text{SD}}$, so we refrain from showing them. The predictions for ID (lower panel) are also well below experimental bounds, see main text for subtleties.
    } \label{fig:mcmc-DM2}
\end{figure}
    We finish the DM discussion by describing some of the necessary details to properly qualify the DD and ID results above. The experimental bounds for the various experiments are indicated by colored curves, see the legends next to the plots, and the experimental bounds themselves sourced from the references listed in Sect.~\ref{sec:computational-setup}. The upper bounds set by various present/past experiments are drawn with a solid line, while the projected sensitivity of DARWIN (with a $200\,\mathrm{t}\,\mathrm{y}$ exposure) and the neutrino fog threshold are drawn with dashed lines. The lines exclude the parameter space above them. The DD plots (upper and middle) show the spin independent (SI) and spin dependent (SD) cross-sections $\sigma_{p}^{\text{SI,SD}}$ of the bino scattering off of protons; scattering off neutrons essentially leads to analogous plots, so we have chosen to omit them from the presentation. For the SI DD case (upper panel),  the most recent LZ results in solid darker green 
    represent the observed $90\,\%$ C.L.~upper exclusion limits on the WIMP-proton cross section, while  
    the green and yellow shaded bands show the expected upper bounds on $\sigma_{p}^{\text{SI}}$ under the background-only hypothesis, with green (yellow) representing the $\pm 1\sigma$ ($\pm 2\sigma$) range, i.e.~the $68\,\%$ ($95\,\%$) range. 
    In the case of ID (lower panel), the quantity of interest is $\langle\sigma v\rangle$, i.e.~the thermally averaged product of cross-section and velocity, for a given annihilation channel. 
    The plot simultaneously presents the results for 
    three such processes: $\tilde{\chi}^{0}_{1}\tilde{\chi}^{0}_{1}\to W^{+}W^{-}$, $\tilde{\chi}^{0}_{1}\tilde{\chi}^{0}_{1}\to ZZ$ and $\tilde{\chi}^{0}_{1}\tilde{\chi}^{0}_{1}\to \tau^{+}\tau^{-}$. For experimental bounds, the type of process is specified in the legend. The grid and MCMC result, however, should be interpreted with care: for each computed parameter point we extract the highest estimate for all three processes. Note that \texttt{MicrOmegas 5.2.7a}  reports the total thermally averaged product of cross-section and relative velocity, and only lists processes with computed branching ratios larger than $10^{-4}$. Since the bino in our scenario is relatively heavy, the branching ratios of the three process of interest are small: DM annihilation into $W^{+}W^{-}$ or $ZZ$ is always below the cutoff value, hence unreported, while annihilation into $\tau^{+}\tau^{-}$ is sometimes above cutoff. For points where none of the three branching ratios are reported, we take their upper bound $10^{-4}$ and don't treat them separately in the statistics; in this sense the plot shows results for tight upper bounds rather than predicted values of $\langle \sigma v\rangle$ for the most most prevalent of the three processes. In light of these subtleties in interpretation, the following conclusion is reaffirmed even more strongly: all processes relevant for ID are well below experimental sensitivity.   
\end{itemize}

\subsection{Viability and testability of scenarios with apparent overabundance of DM \label{sec:tbtau-DM-overabundance}}

As discussed extensively up to now, a neutralino LSP is an attractive candidate for dark matter in supersymmetric models. However, the requirement of reproducing the observed dark matter relic abundance $\OMEGA^0$ places strong constraints on model parameters. Our analysis in Section~\ref{sec:tbtau-results} shows that, within the standard cosmological scenario, the parameter space of our model cannot reproduce the observed relic abundance; instead, it predicts an overabundance, i.e.~$\mathcal{D}>1$. We dedicate this section to a short discussion on the possible origin of the needed dilution after freeze-out and the implications for detecting such scenarios with gravitational waves (GW).

Intriguingly, supersymmetric theories often contain particle species such as the gravitino or moduli (e.g.~the sgoldstino from SUSY breaking) that interact only gravitationally. Due to their extremely suppressed couplings, such particles are expected to decay very late (see, e.g., Refs.~\cite{Ellis:1984eq,Endo:2006zj,Nakamura:2006uc,deCarlos:1993wie,Hasenkamp:2010if,Co:2016xti,Apers:2024ffe}). However, in order not to spoil the successful predictions of standard Big-Bang nucleosynthesis (BBN), these late-decaying moduli must decay before BBN, namely at temperatures $T \gtrsim 5$~MeV~\cite{Moroi:1995fs,Kawasaki:2004qu}.

A long-lived particle that behaves as matter in the early Universe will eventually come to dominate the energy density of the Universe, until it decays at $t\sim \Gamma^{-1}$ ($\Gamma=$ decay width) before the onset of BBN. Through its decay, it injects entropy into the thermal plasma, thereby diluting, for example, the dark matter abundance. The dilution factor $\mathcal{D}$ is then defined as the ratio of the entropy ($s$) before and after its decay, 
\begin{align}
    \mathcal{D}^{-1}= \frac{s_\mathrm{before}}{s_\mathrm{after}} = \frac{\OMEGA^0}{\OMEGA^*},
\end{align}
consistent with the definition of Eq.~\eqref{eq:dilution-factor-definition}, which took an agnostic approach to the dilution mechanism.
We reiterate that $\OMEGA^0$ is the experimentally measured DM relic abundance today, and $\OMEGA^*$ is the DM (over)abundance predicted by our scenario, computed under the assumption of a standard cosmological history. As shown in Fig.~\ref{fig:mcmc1-DM1}, the predicted dilution factor lies in the range $\mathcal{D} \sim \mathcal{O}(10^1) - \mathcal{O}(10^3)$, potentially leading to intriguing cosmological implications, as discussed below.

The $t$-$b$-$\tau$ unification scenario under consideration may originate from an underlying $SO(10)$ gauge symmetry. Within such a setup, depending on the symmetry-breaking chain, promising $SO(10)$ models~\cite{Antusch:2023zjk,Antusch:2024nqg} can produce metastable cosmic strings near the unification scale. The presence of these strings offers a way to test the models, since cosmic strings radiate gravitational waves during the evolution of the universe, which can be detected in ongoing and upcoming experiments. Interestingly, recent pulsar timing array measurements~\cite{Xu:2023wog,Antoniadis:2023ott,NANOGrav:2023gor,Reardon:2023gzh,InternationalPulsarTimingArray:2023mzf} may have observed gravitational waves at very low frequencies of order nano-Hertz, which could point toward metastable cosmic strings with string tensions corresponding to energies close to the GUT scale~\cite{NANOGrav:2023hvm}. It is important to point out that the 68\% credible region~\cite{NANOGrav:2023hvm} in the string tension versus decay parameter plane overlaps with the third Advanced LIGO--Virgo--KAGRA (LVK)~\cite{LIGOScientific:2014pky,VIRGO:2014yos,KAGRA:2018plz} bound, suggesting a degree of dilution in the gravitational wave spectrum during the early universe. Because an early period of matter domination suppresses  the gravitational-wave spectrum at high frequencies (see Ref.~\cite{Antusch:2024ypp} and references therein), it allows the string tension to be raised well above the current LVK limits, as favored by PTA data. In fact, the upcoming Advanced LVK (HLVK) runs have the potential to rule out metastable cosmic string interpretations of the PTA signals unless $\mathcal{D} \gtrsim \mathcal{O}(10)$~\cite{Antusch:2024ypp}, which aligns well with our scenario.

In standard cosmology, the GW spectrum from a cosmic string network exhibits a characteristic flat plateau~\cite{Auclair:2019wcv} at high frequencies. By contrast, the late-time dilution leads to a sharp drop~\cite{Antusch:2024ypp} of the spectrum above a certain frequency, a feature that can be fully probed by upcoming gravitational-wave observatories such as the Laser Interferometer Space Antenna (LISA)~\cite{Audley:2017drz}, Big Bang Observer (BBO)~\cite{Corbin:2005ny}, DECi-hertz Interferometer Gravitational wave Observatory (DECIGO)~\cite{Seto:2001qf}, Einstein Telescope (ET)~\cite{Sathyaprakash:2012jk} and Cosmic Explorer (CE)~\cite{Evans:2016mbw}. Moreover, within our SUSY framework, the additional relativistic species---namely the sparticles---decouple around their respective mass scales, changing the effective number of degrees of freedom and thereby altering~\cite{Antusch:2024ypp} the gravitational-wave spectrum emitted by cosmic strings.
The frequency at which this characteristic change occurs depends on the masses of these particles. As a result, the doubling of particle degrees of freedom in SUSY leaves a detectable imprint on the GW spectrum, which, for our scenario with a SUSY scale of order 10 TeV, could be probed by BBO, DECIGO, ET, and CE.

In summary, we see that the subsequent dilution of $\OMEGA^*$ is physically well-motivated in SUSY scenarios, and that within a presumed $\SO(10)$ GUT setup there may be interesting imprints of this non-standard cosmological history in the stochastic GW background spectrum.

\section{Summary and conclusions \label{sec:conclusions}}
In this study, we have explored the CMSSM parameter space in an $\SO(10)$-inspired scenario of 3rd family Yukawa unification, reparametrized in terms of the ratios \hbox{$x = M_{1/2}/m_0$} and \hbox{$y = A_0/m_0$}. We enforced the correct SM Higgs boson mass $m _{h}$, while leaving the DM relic abundance $\OMEGA^*$ as a prediction, recast as the dilution factor $\mathcal{D}$ from non-standard cosmology needed to explain the measured relic density $\OMEGA^0$.  Given the CMSSM boundary conditions, viable parameter space is found only for $\sgn(\mu)=-1$.

First, the virtues of the $(x,y)$-parametrization are laid out in an example in Section~\ref{sec:grid-example}. By retaining $m_{0}$ as the only parameter with mass dimension, this parameter more directly controls the SUSY scale $Q$ and the SUSY spectrum, and can be determined by setting the SM Higgs boson mass to the experimental value. Furthermore, fixing $\tan\beta$ (or constraining its range), the viable regions of the parameter space become bounded, and a systematic mapping of the viable SUSY scale and neutralino DM realizations across the $(x,y)$ parameter plane is established.

Second, the case of approximate 3rd family Yukawa unification is studied in Section~\ref{sec:tbtau}, by enforcing $t$-$b$-$\tau$ unification to within $5\,\%$ or $10\,\%$ at $\MGUT=2\cdot 10^{16}\,\mathrm{GeV}$. This indirectly limits the range for $\tan\beta$ and thus 
again provides a bounded region in the transformed CMSSM parameters $\{m_{0},x,y,\tan\beta\}$. Our analysis, carried out through both grid-based scans and MCMC techniques, reveals a highly constrained scenario, predicting the following: 
\begin{enumerate}[label=(\roman*),itemsep=-0.1cm,topsep=0.2cm]
    \item a predicted spectrum of SUSY particles with a SUSY mass scale of $\mathcal{O}(10)\,\mathrm{TeV}$, featuring
    \item  heavy extra Higgs partners $H,A,H^\pm$ at $\sim 2\div 3\,\mathrm{TeV}$, just above current LHC bounds and within reach of HL-LHC,
    \item a bino dark matter candidate leading to overabundance, necessitating a dilution factor of $\mathcal{D}=\mathcal{O}(10^{1\div 3})$ from non-standard cosmology, with interesting potential signatures in the stochastic GW background.  
\end{enumerate}
Since predictive $\SO(10)$ models can give rise to cosmic strings, intriguingly, the required dilution factor in our setup---typically arising from late-decaying gravitino or moduli (e.g.~sgoldstino) in local SUSY frameworks---can leave characteristic imprints in gravitational wave observations, potentially testable in upcoming detectors such as LISA, ET, and CS. 

In summary, the presented framework thus provides a coherent way to connect Grand Unification-based Yukawa relations with predictions for the supersymmetric mass spectrum, LHC-accessible cousins of the SM Higgs, and dark matter phenomenology. The approximate agreement between grid and MCMC results further motivates the use of MCMC techniques for future explorations of more realistic GUT boundary conditions with enlarged parameter space compared to the CMSSM. Finally, the recast parameterization we employed offers a transparent perspective on how these elements align, and it can serve as a useful basis for further theoretical investigations and phenomenological tests in the context of the CMSSM and beyond.


\section*{Acknowledgments}
We thank Emanuele Bagnaschi for a useful discussion on FeynHiggs. 
SS acknowledges the financial support
from the Slovenian Research Agency (research core funding No.~P1-0035 and N1-0321).
VS is supported by the European Union---Next Generation EU and
by the Italian Ministry of University and Research (MUR) 
via the PRIN 2022 project n.~2022K4B58X---AxionOrigins.

\bibliographystyle{style.bst}
\providecommand{\href}[2]{#2}\begingroup\raggedright\endgroup

\end{document}